\documentclass{emulateapj}
\slugcomment{{\sc Accepted to ApJ:} October 23rd, 2013}

\usepackage{epsfig}
\usepackage{graphicx}
\usepackage{float}

\newcommand{\lambdare}{$\lambda_{Re}$}
\newcommand{\lambdar}{$\lambda_{R}$}
\newcommand{\atlas}{ATLAS\textsuperscript{3D}}

\begin{document}
\title{Angular Momenta, Dynamical Masses, and Mergers of Brightest Cluster Galaxies \footnotemark[*]}
\author{Jimmy\footnotemark[1], Kim-Vy Tran\footnotemark[1], Sarah Brough\footnotemark[2], Karl Gebhardt\footnotemark[3], Anja von der Linden\footnotemark[4]\textsuperscript{,}\footnotemark[5]\textsuperscript{,}\footnotemark[6], Warrick J. Couch\footnotemark[7], Rob Sharp\footnotemark[8]}
\footnotetext[*]{Based on VLT service mode observations (Programs 381.B- 0728 and 087.B-0366) gathered at the European Southern Observatory, Chile.}
\footnotetext[1]{ George P. and Cynthia W. Mitchell Institute for
Fundamental Physics and Astronomy, Department of Physics and
Astronomy, Texas A\&M University, College Station, TX 77843, USA }
\footnotetext[2]{Australian Astronomical Observatory, PO Box 915, North Ryde, NSW 1670, Australia}
\footnotetext[3]{Department of Astronomy, University of Texas at Austin, 1 University Station C1400, Austin, TX 78712, USA}
\footnotetext[4]{Kavli Institute for Particle Astrophysics and Cosmology, Stanford University, 452 Lomita Mall, Stanford, CA 94305-4085, USA }
\footnotetext[5]{Department of Physics, Stanford University, 382 Via Pueblo Mall, Stanford, CA 94305-4060, USA}
\footnotetext[6]{Dark Cosmology Centre, Niels Bohr Institute, University of Copenhagen, Juliane Maries Vej 30, 2100 Copenhagen \O,
Denmark}
\footnotetext[7]{Centre for Astrophysics and Supercomputing, Swinburne University, PO Box 218, Hawthorn, VIC 3122, Australia}
\footnotetext[8]{Research School of Astronomy \& Astrophysics, Australian National University, Cotter Road, Weston Creek, ACT 2611, Australia}

\begin{abstract}
Using the VIMOS Integral Field Unit (IFU) spectrograph on the Very Large Telescope (VLT), we have spatially mapped the kinematic properties of 10 nearby Brightest Cluster Galaxies (BCGs) and 4 BCG companion galaxies located within a redshift of $z=0.1$. In the hierarchical formation model, these massive galaxies $(10^{10.5} M_{\sun} < M_{dyn} < 10^{11.9} M_{\sun})$ are expected to undergo more mergers than lower mass galaxies, and simulations show that dry minor mergers can remove angular momentum.  We test whether BCGs have low angular momenta by using the $\lambda_{Re}$ parameter developed by the SAURON and ATLAS\textsuperscript{3D} teams and combine our kinematics with Sloan Digital Sky Survey (SDSS) photometry to analyze the BCGs' merger status.  We find that 30\% (3/10) of the BCGs and 100\% of the companion galaxies (4/4) are fast rotators as defined by the ATLAS\textsuperscript{3D} criteria.  Our fastest rotating BCG has a $\lambda_{Re}=0.35\pm0.05$.  We increase the number of BCGs analyzed from 1 in the combined SAURON and ATLAS\textsuperscript{3D} surveys to 11 BCGs total and find that above $M_{dyn}\sim11.5 M_{\sun}$, virtually all galaxies regardless of environment are slow rotators.  To search for signs of recent merging, we analyze the photometry of each system and use the $G-M_{20}$ selection criteria to identify mergers.  We find that $40\pm20$\% of our BCGs are currently undergoing or have recently undergone a merger (within 0.2 Gyrs).  Surprisingly, we find no correlation between galaxies with high angular momentum and morphological signatures of merging.
\\

\end{abstract}

\maketitle

\section{Introduction}

Galaxies are expected to grow hierarchically in a $\Lambda$CDM universe.  In this framework, clouds of dark matter cool down and begin to collapse, forming the structure upon which galaxies and clusters are formed (\citealt{Peebles:69}, \citealt{Doroshkevich:70}, \citealt{White:84}).  Simulations of mergers of dark-matter halos have shown that major mergers (progenitor mass ratios of 1:1 or 2:1) increase the angular momentum of the combined dark-matter halo \citep{Vitvitska:02}.  Other studies have shown that minor mergers either slightly increase or simply preserve the angular momentum of a dark-matter halo \citep{D'Onghia:04}.  Without taking into account the interactions of baryons, merger events (whether major or minor) appear to increase a dark-matter halo's specific angular momentum.

When accounting for baryons, the connection between mergers and angular momentum changes slightly.  Although the effects of baryons within 1 effective radius ($R_e$) are uncertain (\citealt{vandenBosch:02}, \citealt{deJong:04}), we can examine merger simulations that include baryons to see how merging galaxies behave on a large scale.  Spiral galaxies, which have a high initial angular momentum, have cold stellar discs that are fragile and easily destroyed in mergers \citep{Naab:06a}.  When these initially discy high angular momentum galaxies merge together, they form early-type galaxies that were expected to be dispersion-supported and not rotation-supported.  However \citet{Emsellem:07} showed that early-type galaxies can still exhibit a high angular momentum, and in fact the majority of them do (86\% in \citealt{Emsellem:11}).  The mechanism by which these early-type galaxies have retained or regained their angular momentum is still uncertain.

One possible explanation for the high angular momentum observed in early-type galaxies comes from simulations of gas-rich mergers.  In the early universe ($z > 2$), before feedback mechanisms have had a significant impact on galaxy composition, galaxy mergers are expected to be gas-rich \citep{Tacconi:10}.  Simulations of mergers of equal-mass, gas-rich galaxies (wet major mergers) have been shown to produce high angular momentum galaxies \citep{Naab:06a}.  The dissipational effect of gas appears to be an important factor in forming a high angular momentum merger remnant (\citealt{Bender:92}, \citealt{Kormendy:96}, \citealt{Faber:97}, \citealt{Naab:06a}).

However gas-rich mergers are only expected at intermediate/high redshifts.  From $z=2$ until the present, as feedback mechanisms remove gas from the galaxies, mergers are expected to be increasingly gas-poor and dissipationless.  In the case of major gas-poor (dry) mergers, \citet{Bois:11} have shown that major dry mergers are capable of forming galaxies with a variety of angular momenta, depending upon the initial orbital parameters.  They find little dependence of a merger remnants angular momentum on the progenitor galaxy's gas fraction.  For a galaxy to transform from an initially fast rotating galaxy into a slowly rotating galaxy, said galaxy has to accrete at least half of its mass via dry minor mergers (\citealt{Bournaud:07}, \citealt{Jesseit:09}, \citealt{Bois:10}, \citealt{Bois:11}).  The environment in which a galaxy exists will determine the frequency of major and minor galaxy mergers.

\subsection{Merger History of Brightest Cluster Galaxies}

Brightest Cluster Galaxies (BCGs) exist in an extreme environment.  They are typically located in the center of their cluster, near the bottom of the cluster's potential well.  As a result of their location, BCGs are expected to undergo more mergers than a typical galaxy.  The increased frequency of merging events in BCGs should make them sensitive to the relation between galaxy mergers and angular momentum.

The large number of merging events is also likely the cause of the high mass and early-type classification observed in a typical BCG.  Observations by \citet{Lidman:12} have shown that BCGs have grown in mass since $z=1.6$ and on average, have grow by a factor of $1.8 \pm 0.3$ from $z=0.9$ to $z=0.2$.  Similarly, simulations of galaxy growth via mergers by \citet{DeLucia:07} have shown that only half the mass of a BCG is in place by $z=0.5$.  The same simulations found that the majority of stars within a BCG formed very early, with 50\% of the stars being formed before $z=5$ \citep{DeLucia:07}.  In this scenario, BCGs would have accumulated half of their mass in the time frame from $z=0.5$ to $z=0$ via gas-poor mergers.

The fact that multiple minor dry mergers have been shown to remove angular momentum from a galaxy when the mergers are isotropically distributed, combined with the fact that BCGs in simulations undergo a higher than average number of dry merger events, leads us to hypothesize that BCGs are likely to have a lower average angular momentum when compared to standard elliptical galaxies.  In other studies BCGs have been shown to be remarkably uniform and exhibit special characteristics when compared to elliptical galaxies not at the center of their cluster.  \citet{vonderLinden:07}, showed that a typical BCG is more likely to host a radio-loud AGN, is larger in size, and has higher velocity dispersion on average than a typical elliptical galaxy in the same mass range.  These currently observed differences are likely the result of the combination of two effects: BCGs' large stellar masses and BCGs' location at the bottom of their host clusters' potential well , although it is uncertain which effect dominates.

Specific features of BCGs, such as their large radii and low surface brightnesses compared to normal elliptical galaxies are consistent with the products of major dissipationless mergers (e.g. \citealt{Oegerle:91, Brough:05, vonderLinden:07, Lauer:07, Tran:08}).  Their sizes and velocity dispersions may have also evolved faster than less-massive early-type galaxies since $z \sim 0.3$ (\citealt{Bernardi:09} although, c.f. \citealt{Stott:11}).  These two factors cause the Faber-Jackson relation between luminosity and velocity dispersion to be flatter for BCGs.  Similar studies of the Fundamental Plane (FP) have found differing slopes between populations of BCGs and elliptical galaxies \citep{Desroches:07}.  The differing slope of the Faber-Jackson relationship and the Fundamental Plane for BCGs compared to ellipticals supports the scenario where BCGs form mainly via dissipationless mergers \citep{Boylan-Kolchin:08}.

\subsection{ Signatures of Recent Merging in BCGs at $z<0.1$ }

By combining spectroscopic information with photometric information, we examine the recent merger history of a sample of 10 BCGs.  Integral Field Unit (IFU) spectroscopy allows us to observe the kinematic properties of galaxies in two dimensions, showing whether they are largely rotation-supported or dispersion-supported.  The SAURON team have developed the \lambdar\ parameter, which utilizes the increased spatial information from IFU spectroscopy to quantify the observed stellar angular momentum in galaxies \citep{Emsellem:07}.  They used this parameter to quantify the angular momentum of 48 early-type galaxies.  This work was followed by the ATLAS\textsuperscript{3D} team, which performed a similar analysis on 260 galaxies \citep{Emsellem:11}.  In both studies they classified galaxies as either Fast Rotators (FR) or Slow Rotators (SR) based on their \lambdar\ value within $1 R_e$.

The \lambdar\ parameter provides a quantitative method to compare the stellar kinematics of BCGs to those of other early-type galaxies, however the SAURON sample has only 4 galaxies with $M_{dyn} > 10^{11.5} M_\sun$ and only includes one BCG (M87).  The ATLAS\textsuperscript{3D} survey studies an additional 10 galaxies with $M_{dyn} > 10^{11.5} M_\sun$.  Above $10^{11.5} M_\sun$ only 23\% of galaxies from the ATLAS\textsuperscript{3D} survey are classified as fast rotators \citep{Emsellem:11}.  The mass limit of $10^{11.5}$ may be important as \citet{Peng:10} have shown that above $M_{*}  = 10^{11.5} M_\sun$ the majority of galaxies have undergone a major merger after their star formation was quenched.  If dry mergers preferentially remove angular momentum, then we would expect to see fewer BCGs above $10^{11.5} M_\sun$ with a high \lambdar\ value.

In order to determine the recent merger history of our galaxies, we use a combination of Source Extractor (SExtractor) \citep{Bertin:96}, GALFIT \citep{Peng:02}, and PyMorph \citep{Vikram:10} to produce photometric models of these galaxies.  We then subtract our models from the original image to identify tidal tails, multiple cores, and excess intra-cluster light, which would all be signs of merging that can be identified visually.  A more quantitative detection of merging comes from the $G - M_{20}$ (Gini coefficient minus the 2nd order moment of the brightest 20\% of pixels) value for each galaxy.  According to simulations, these parameters would tell us if there was a dry merger within the last 0.2 Gyrs \citep{Lotz:11}.  After 0.2 Gyrs merger information will have been erased by dynamical friction, meaning this selection criterion is only able to inform us of very recent or currently ongoing mergers.

It is our goal to measure \lambdar , dynamical mass, and recent merger history for our sample of BCGs and compare our kinematic results to the SAURON and ATLAS\textsuperscript{3D} results for early-type galaxies.  We present targeted observations of BCG stellar kinematics and photometrics for 7 BCGs with close companions and 3 BCGs with no close companion.  In our initial selection, we specifically chose BCGs with close companions to determine if the companions also have high \lambdar .  We also determine if morphological signatures of merging are correlated with \lambdare\ measurements.

We assume a Hubble constant of H$_0$ = 70 km s$^{-1}$ Mpc$^{-1}$ and $\Omega_M$ = 0.3, $\Omega_\Lambda$ = 0.7.

\vspace{10 mm}

\section{Observations}

\begin{deluxetable*}{lrrrrrrr} [h]
\tablecolumns{8}
\tablecaption{IFU Observational properties of BCGs and companions.}
\startdata
\hline
\hline
	Galaxy & RA & Dec & Seeing ($^{\prime\prime}$) & Integration & Number of & Observation Date & Cluster Velocity\\
	 (SDSS-C4-DR3) & (J2000) & (J2000) & $\pm$ 0.01 & Time (s) & exposures &  & Dispersion $\sigma$ (km s$^{-1}$)\\
	 
\hline
\\
1027 & 12:47:43.4 & -00:09:07 & 1.0 & 6900 & 6 & 04-05-08 \& 06-03-08 & 1020\\
1042 & 15:15:18.0 & +04:22:54 & 0.8 & 6000 & 6 & 05-04-11 \& 05-30-11 & 857\\
1048 & 13:42:09.6 & +02:13:38 & 1.0 & 6000 & 6 & 04-08-11 \& 07-30-11 & 828\\ 
1050 & 13:44:25.8 & +02:06:36 & 0.8 & 3450 & 3 & 04-27-08 & 514\\
1066 & 13:31:10.8 & -01:43:49 & 0.8 & 3450 & 3 & 05-12-08 & 814\\
1153 & 16:04:13.7 & +00:03:13 & 0.8 & 6800 & 6 & 04-05-08 \& 06-04-08 & 295\\
1261 & 12:25:33.4 & +09:23:29 & 0.8 & 6000 & 6 & 04-30-11 & 520\\
2001 & 23:24:18.0 & +14:38:00 & 0.9 & 3450 & 3 & 08-09-08 & 695\\
2039 & 22:31:43.2 & -08:24:32 & 1.2 & 10000 & 10 & 06-040\-11 \& 06-05-11 & 505\\
 & & & & & & 06-29-11 \& 06-30-11 &\\
2086 & 23:22:56.4 & -10:02:44 & 1.2 & 3450 & 3 & 08-02-08 & 599\\
\enddata
\tablecomments{Seeing is the average over all exposures for each galaxy.  BCGs 1027, 1050, 1066, and 2086 were also analyzed in \citet{Brough:11}.  Cluster velocity dispersion is form \citet{vonderLinden:07}}
\label{obs_table}
\end{deluxetable*}

\subsection{Spectroscopic Measurements}

In \citet{vonderLinden:07}, a sample of 625 BCGs ($z < 0.1$) were selected from the C4 cluster catalogue \citep{Miller:05} of the Third Data Release of the Sloan Digital Sky Survey (SDSS; \citealt{York:00}).  In \citet{Brough:11} BCGs with companions within $\sim$ 10$^{\prime\prime}$ (18 kpc at $z \sim 0.1$) were chosen from the von der Linden sample.  We are using the same 4 galaxies from \citet{Brough:11} as well as 7 new galaxies chosen by the same criteria as \citet{Brough:11}.  The redshifts of these objects are in the range $0.04 < z < 0.1$.  As a matter of shorthand, we will drop the SDSS-C4-DR3 prefix from all target BCG names, and simply use the last 4 digits that are unique to each cluster.  In total, our initial sample size is 7 BCGs with companions and 4 BCGs without companions within $\sim 10^{\prime\prime}$.  One target, BCG 1067, will be omitted from analysis because it had excessively noisy observations and we were unable to obtain results on that BCG above our signal-to-noise cut of 10.

The first set of BCGs were observed with VIMOS \citep{LeFevre:03} on the VLT from April to August of 2008 (Prog. ID 381.B-0728) and the second set were observed, also with VIMOS, from April to July of 2011 (Prog. ID 087.B-0366). VIMOS was used in IFU mode with the high-resolution blue grism and a spatial sampling of $0.67^{\prime\prime}$/pixel (BCGs 1153 and 1067 were observed with a spatial sampling of $0.33^{\prime\prime}$/pixel). This gives a field-of-view (FOV) of $27^{\prime\prime}$ x $27^{\prime\prime}$ ($13^{\prime\prime}$ x $13^{\prime\prime}$ for 1153 and 1067).  The VIMOS HR Blue grism (pre March 15th 2012 version) has a spectral range of 4150 - 6200 \AA\ and a spectral resolution of 0.51 \AA\//pixel.  Observations were made during dark time, with an average seeing of $0.9^{\prime\prime}$.  Average seeing for individual galaxies can be seen in Table \ref{obs_table}.

\subsubsection{IFU Data Reduction}

Initial IFU data reduction is achieved using the VIMOS Pipeline \citep{Izzo:04}.  A VIMOS field-of-view is split into four quadrants, and initially each quadrant is reduced separately by the VIMOS Pipeline.  The VIMOS Pipeline makes the necessary calibration files (master bias, arc spectrum, flat field, fiber identification, etc.).  The VIMOS Pipeline also extracts the science spectrum in each fiber from the raw science frames.

The science spectra are then fed into IDL routines, where we first mask the bad fibers so that they do not interfere with the data reduction process.  For each quadrant, a sky background spectrum is calculated by taking the median over the spaxels (spatial pixels) that do not contain a significant contribution of light from the target galaxy.  Sky spaxels are selected by plotting the intensity in every spaxel, and looking for a sustained minimum in the intensities across several adjacent fibers.   We perform a Gaussian fit to the 5577 \AA\ [OI] skyline.  This fit is used to normalize the transmission in each spaxel across the whole quadrant.  Since the strength of the skyline should be the same in all spaxels, each individual spectra is scaled so that its skyline flux matches the median skyline flux across all spaxels. We then subtract the Gaussian fit from the spectra to remove the skylines.  Once this has been completed for each quadrant, we combine the four quadrants into one data cube, and then renormalize transmission across the whole cube using the strength of the skylines once again (skyline information was preserved from the earlier step).  This provides us with a data cube containing two dimensional spatial information, with the spectrum being the third dimension.

Each BCG observation consists of multiple dithered exposures, so we stack the multiple exposures with a $5\sigma$ clipped mean.  With our stacked data cube, we then perform a signal-to-noise cut, where we discard any spaxel with an overall S/N ratio less than 5.  We use the 2-dimensional adaptive spatial binning code of \citet{Cappellari:03} to re-bin pixels to a minimum S/N ratio of 10 per pixel.  This takes the fibers that are below the final S/N cutoff and finds fibers near them to bin together so they reach our required S/N ratio.

\subsubsection{Stellar Kinematics}

The stellar kinematics (velocity, V, and line-of-sight velocity dispersion, $\sigma$) of the galaxy are computed from the binned spectra using a penalized pixel fitting scheme, (pPXF; \citealt{Cappellari:04}) and the MILES (Medium-resolution Isaac Newton Telescope Library of Empirical Spectra; \citealt{Sanchez-Blazquez:06}) evolutionary stellar population templates. This method determines the stellar kinematics by fitting templates to the absorption line features in the spectra.  We choose 10 templates of stars in the range of G6 to M1 (luminosity classes III, IV, and V) from the MILES library and convolve them using the quadratic difference between the resolution of the library and the observations.  The pPXF fits are performed in the region around the observed G-band and Calcium H and K absorption features (4200 - 4900 \AA).  The MILES templates cover a similar wavelength range to VIMOS and have a similar spectral sampling (FWHM = 1.8 \AA). 

\begin{figure} [h]
\epsfig{ file=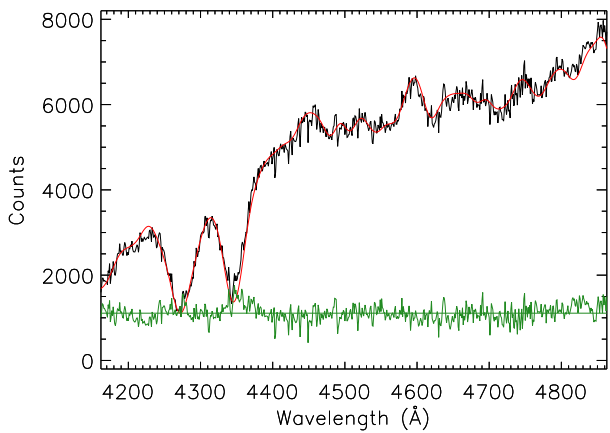, scale=0.5 }
\caption{An example of a fit produced by pPXF.  The spectrum pictured here is the result of co-adding every spaxel within the effective radius of the BCG 1050.  The black line is the science spectrum, and the red line is the best fit produced by pPXF.  The green line is the residual from the best fit.}
\label{example_spectra}
\end{figure}

Velocity dispersion for a galaxy as a whole is found by co-adding all fibers within the effective radius of a galaxy into one bin, and performing a pPXF fit on the binned spectrum as in \citet{Cappellari:06}.  As an alternative method for calculating the velocity dispersion of the whole galaxy, we also tried calculating the luminosity weighted mean of each individual bin's velocity dispersion.  These results were often not in agreement with the effective spectrum method.  However, by integrating (preliminary) inclination information as described in Eq. 29 of \citet{Cappellari:13}, the two velocity dispersion measuring methods appear to be in agreement.  We choose to use the effective spectrum method instead of the weighted mean method because of the higher S/N ratio afforded by binning many spaxels together.  The effective spectrum method also simulates an observation with a single aperture of radius $R_e$.

We perform 100 Monte Carlo realizations on each observation to estimate the uncertainty in our results.  This process involves adding random noise on the order of the background noise to each binned spectrum, and performing the pPXF fit again on the noisy data.  We repeat this process 100 times, then take a weighted mean of all the fits with the random noise added.  The standard deviation of all the fits is our uncertainty for our velocity and velocity dispersion results.  Typical uncertainties of both values are on the order of 20 km/s.  Errors on $M_{dyn}$ and \lambdar\ are propagated from these errors using standard Taylor series error propagation techniques.

The IDL code described above to reduce our data is publicly available\footnotemark[1].
\footnotetext[1]{http://galaxies.physics.tamu.edu/index.php/Jimmy\#Code}

\subsection{Photometric Measurements}

Photometric observations are taken from SDSS Data Release 3.  For our analysis we use only the r-band images of each galaxy.  Each galaxy is modeled twice, once using a combination of Source Extractor \citep{Bertin:96} and GALFIT \citep{Peng:02} with a single component de Vaucouleurs profile, and a second time using PyMorph \citep{Vikram:10} to obtain the Gini and $M_{20}$ coefficients.   The de Vaucouleurs profile is used to find the effective radius of each galaxy.

Effective radius measurements are found by first feeding the SDSS r-band image into Source Extractor to obtain a catalog of all sources identified in the image.  We use the objects catalog to determine the centers of the galaxies we wish to model in GALFIT.  We also invert the segmentation map produced by Source Extractor (non-zero pixels are set to zero, and pixels that were originally zero are set to one) to generate the bad pixel mask that will be used with GALFIT.  The default values and keyword searches were used as inputs to Source Extractor except the following: MAG\_ZEROPOINT = 25.61, BACK\_SIZE = 256, and PIXEL\_SCALE = 0.396.  The original SDSS r-band images are then fed into GALFIT using the position, magnitude, effective radius, axis ratio, and position angle found by Source Extractor as initial assumptions.  We model each galaxy using a single component de Vaucouleurs profile in GALFIT.  We generate our Point Spread Function (PSF) by using a 2-4 gaussian fit in GALFIT to model a point source near each galaxy.  Residual maps are inspected by eye to determine if the fit appears to be reasonable, and if not, initial parameters are adjusted until a reasonable fit is found.

PyMorph combines Source Extractor, GALFIT, and other added analysis routines in a python wrapper in order to automate the photometric analysis process.  It is used to determine the merger properties of our sample of galaxies.  We use it to make two component models for each galaxy.  We then perform a visual inspection of the residual image to check for the goodness of the fit.

PyMorph also reports the Concentration, Asymmetry, Clumpiness, Gini Coefficient, and $M_{20}$ for each galaxy that is modeled.  We use the relation between the Gini Coefficient and $M_{20}$ to search for recent mergers.  The Gini coefficient measures uniformity of the distribution of a galaxyÕs light \citep{Abraham:03} and $M_{20}$ measures the distribution of the brightest 20\% of pixels.  The relation between these two properties tells us if there are irregularities in the galaxy's light distribution, signaling some type of violent interaction.  We can rule out the possibility that anomalies in these parameters are due to projection effects because we can measure the redshift of each galaxy to confirm that they are indeed located near each other in physical space.

\vspace{10 mm}

\section{Results}

\subsection{Kinematic Maps}

\begin{figure}
\vspace{-3 mm}
\epsfig{ file=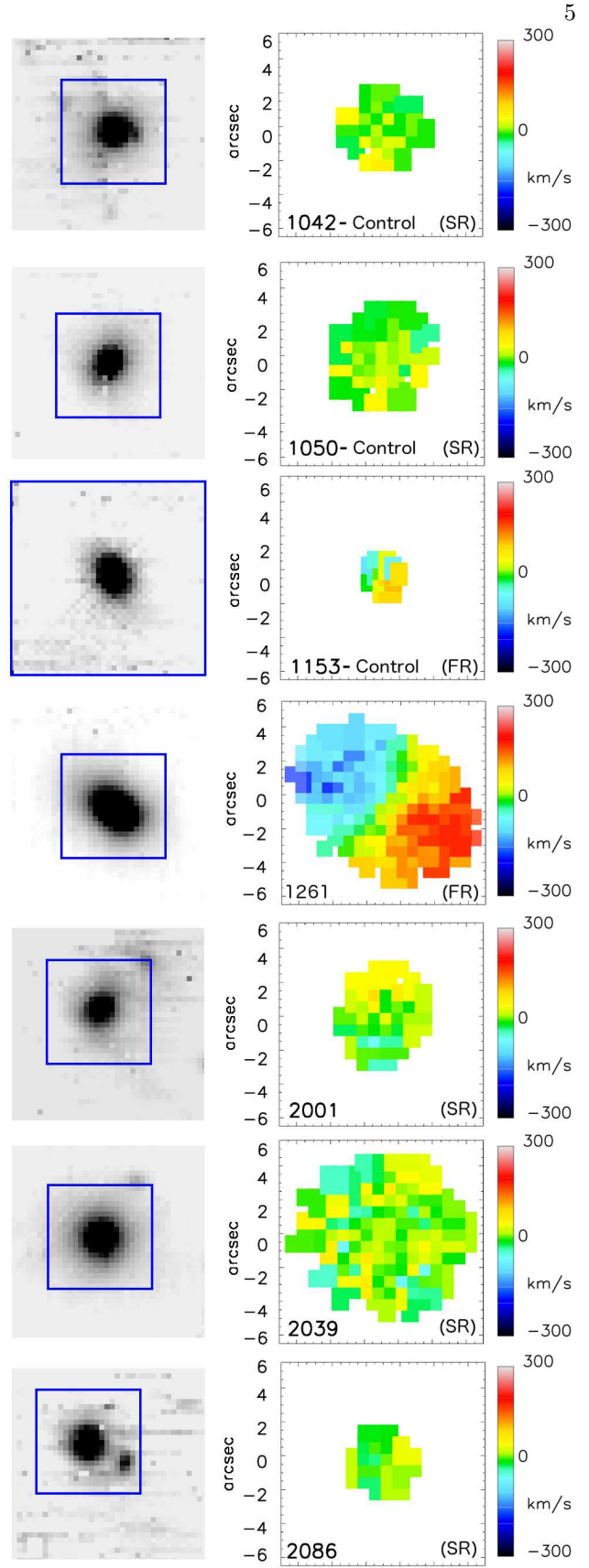, scale = 1.0}
\caption{Left: Collapsed IFU images showing the full VIMOS FOV with the $12^{\prime\prime}$ x $12^{\prime\prime}$ boundary of the velocity maps highlighted in blue.  Right: Velocity maps of BCGs scaled to $12^{\prime\prime}$ x $12^{\prime\prime}$.  Only spaxels with a S/N $> 10$ are shown.  BCGs 1050 (SR),  1042 (SR), and 1153 (FR) were chosen as control galaxies with no companions within $10^{\prime\prime}$.  Slow rotating galaxies (1042, 1050, 2001, 2039, 2086) exhibit very little change in velocity across the galaxy, suggesting that these are either completely face-on or dispersion-supported galaxies.  In the fast rotating galaxies (1261, 1153) there is a velocity gradient across the major axis, showing evidence of rotation.}
\label{single_velocities}
\end{figure}

\begin{figure*}
\epsfig{ file=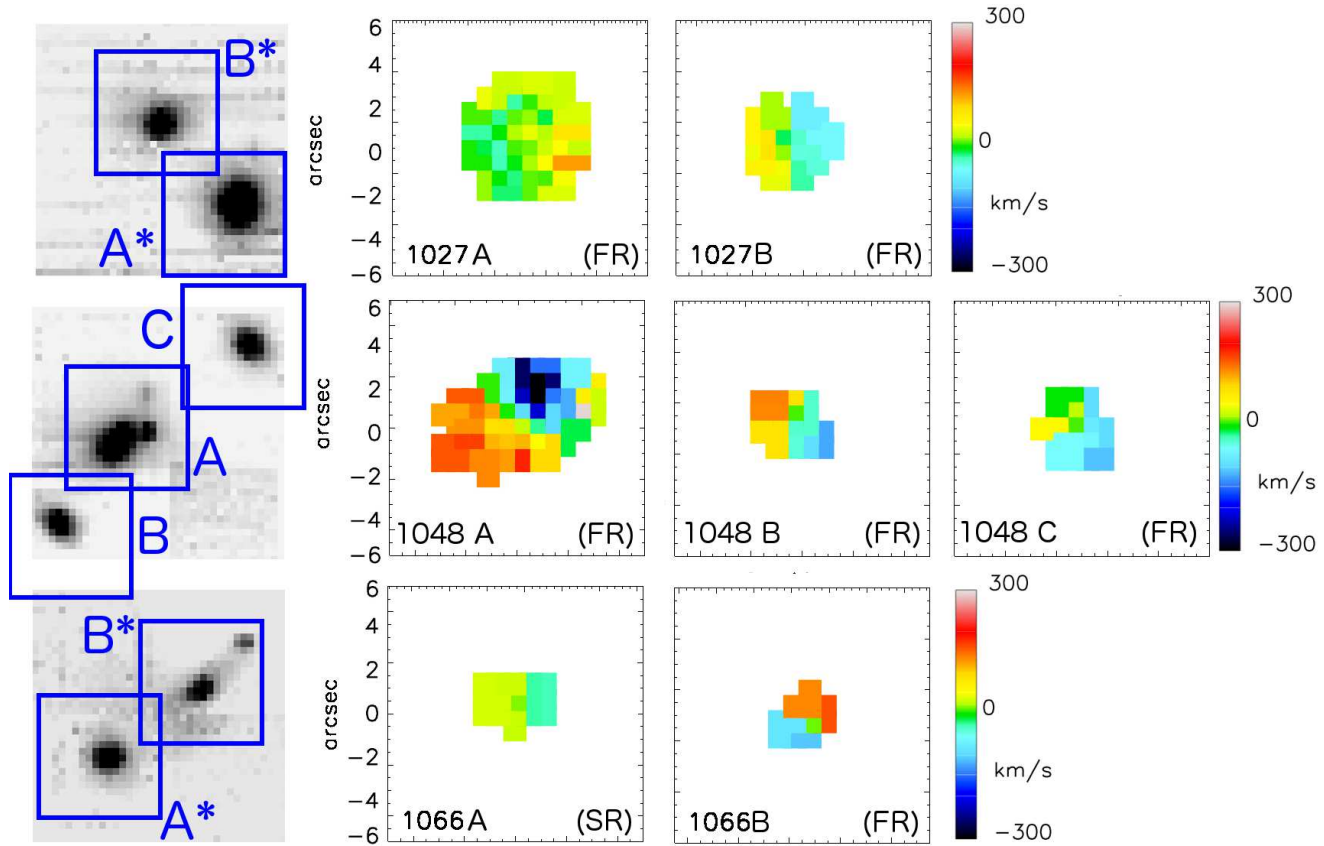, scale = 1.2}
\caption{Left: Collapsed IFU images showing the full VIMOS FOV with the $12^{\prime\prime}$ x $12^{\prime\prime}$ boundary of the velocity maps highlighted in blue.  Right: Velocity maps of BCGs scaled to $12^{\prime\prime}$ x $12^{\prime\prime}$.  BCGs are labeled with the letter A, the brightest companion is labeled B, and so on.  BCG 1048, as well as the companions of the BCGs in clusters 1027, 1048, and 1066 all show signs of rotation.  Both companions of 1066 and 1027 were shown to be bound to their BCG neighbor in \citet{Brough:11} However neither outer companions are gravitationally bound to BCG 1048.  Bound systems are indicated with a star in collapsed IFU images.  Velocities shown for each galaxy are relative to the individual galaxy's redshift, and not absolute across the whole system.}
\label{multi_velocities}
\end{figure*}

We begin our spectroscopic analysis with the velocity maps seen in Figures \ref{single_velocities} \& \ref{multi_velocities}.  Our first goal is to search for visual signs of galaxy rotation.  Figure \ref{single_velocities} shows the velocity maps of the galaxies without companion measurements. This sample contains three of our control galaxies, BCGs 1042, 1050, and 1153, which were selected to not have any companions within $10^{\prime\prime}$.  The S/N on the companion galaxies of BCGs 2001, 2039, and 2086 was below 10 after binning, meaning we were unable to achieve accurate measurements of the companion's kinematic properties.  

By visual inspection of Figure \ref{single_velocities}, we see that 5 of the BCGs (1042, 1050, 2001, 2039, and 2086) appear to have little to no rotation.  They have a uniform velocity across the entire BCG and the velocities measured in each spaxel of the BCG are very near 0 km/s.  These observations are consistent with a dispersion-supported early-type galaxy.  Two of the BCGs (1153 and 1261) in Figure \ref{single_velocities} appear to have velocity gradients across their surface, with BCG 1261 being the most obvious example.

\begin{figure}
\hspace{-7 mm}\epsfig{ file=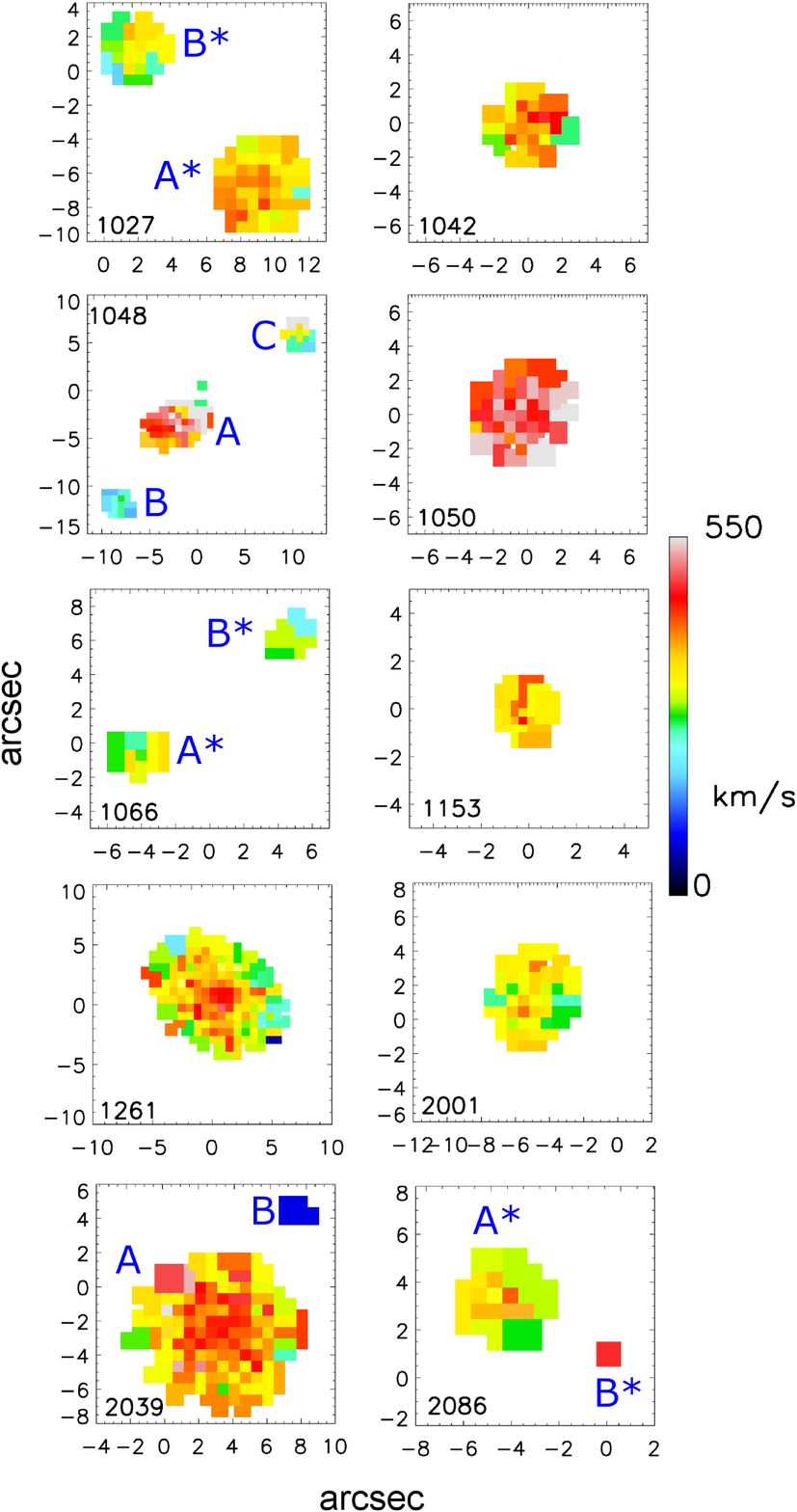, scale=0.2}
\caption{Velocity dispersion maps of each galaxy.  BCGs are labeled with the letter A, the brightest companion is labeled B, and so on.  Bound systems are indicated with a star.  In BCG 1261 there is a rise in the dispersion in the center of the BCG, suggesting that the velocity gradient seen in the velocity maps is due to rotation.}
\label{dispersions}
\end{figure}

Figure \ref{multi_velocities} shows velocity maps of the BCGs that have bright enough companions to measure the companion's kinematic properties.  In each collapsed IFU field-of-view, the BCG is labeled with the letter A, and then brightest companion is labeled B, and so on.  BCG 1048 appears to have one side very red-shifted and the opposite side very blue-shifted, suggesting rotation.  Note also that all the companions in Figure \ref{multi_velocities} appear to exhibit rotation.

The velocity dispersion map (Figure \ref{dispersions}) shows very high dispersions in BCG 1048, suggesting a possible ongoing interaction.  We also see in Figure \ref{dispersions} a peak in the dispersion near the center of BCG 1261.  This peak could be the result of seeing both the positive and negative velocities on either side of the axis of rotation in the same fiber, enhancing the observed dispersion.

\subsection{Angular Momenta (\lambdar )}
\label{section:lr}

\begin{figure}
\hspace{ -5 mm }\epsfig{ file=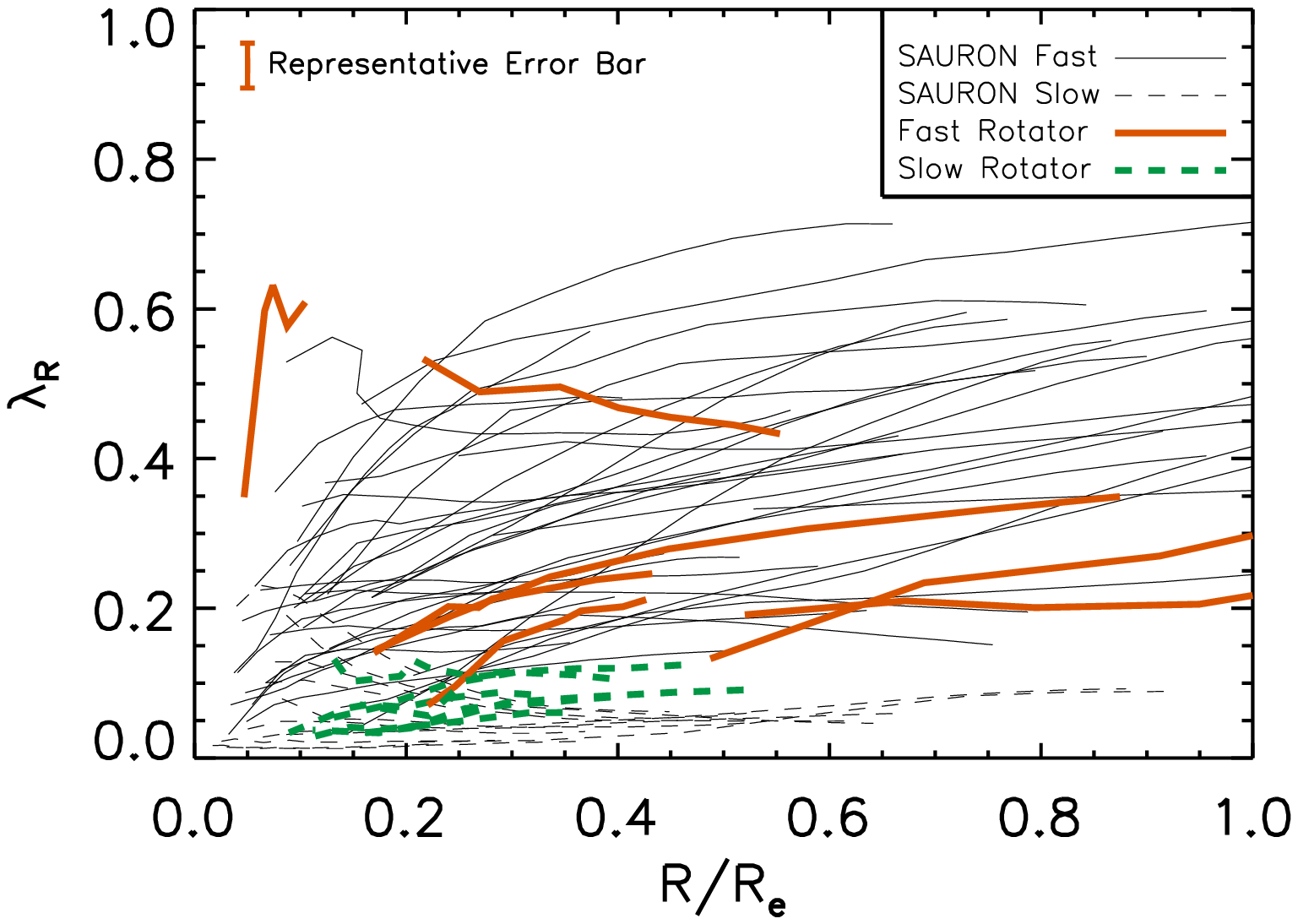, scale=0.5 }
\caption{Angular momentum profile.  Our sample of galaxies are plotted with thick orange and green lines.  The SAURON sample is plotted in grey.  We choose not to plot the full ATLAS\textsuperscript{3D} sample of 260 galaxies in order to keep the plot legible.  Fast rotators are plotted as solid lines and slow rotators are plotted as dashed lines.  Fast rotators have a convex profile, and slow rotators have a concave profile.}
\label{lambda_v_rad}
\end{figure}

In order to quantify the rotation seen in some of the BCGs mentioned above, we use the \lambdar\ parameter developed by the SAURON team \citep{Emsellem:07}.  This parameter acts as a proxy for angular momentum, and is defined as
\begin{equation} \lambda_{R} \equiv \frac{\langle R | V | \rangle}{\langle R \sqrt{V^2+\sigma^2} \rangle} \label{lambdar} \end{equation}
where R is the distance of the spaxel to the galaxy center, V is the velocity of the spaxel, and $\sigma$ is the velocity dispersion.  The numerator acts as a surrogate for the angular momentum L, and the denominator acts as a mass normalization.  The brackets in the numerator and the denominator denote a luminosity weighted average.

The \lambdar\ profile for each BCG and companion galaxy, plotted along with the SAURON results, can be seen in Figure \ref{lambda_v_rad}.  A higher \lambdar\ value indicates higher angular momentum.  As expected angular momentum tends to increase with radius, especially in galaxies classified as fast rotators.  Galaxies that fit into the Fast Rotator (FR) category appear to have a convex profile as \lambdar\ increases with radius, whereas most Slow Rotator (SR) galaxies have a concave profile.  Most of our BCGs appear to have profiles consistent with the slow rotator category.

\lambdare\ is the measured angular momentum at the effective radius.  In cases where S/N of our measurements drops below 10 before we reach $1 R_e$, we assume that value of \lambdare\ is the furthest measured \lambdar\ in that galaxy.  This likely gives a minimum value for \lambdare\ because \lambdare\ tends to increase with radius within $1 R_e$ as observed in the SAURON results in Figure \ref{lambda_v_rad}.

The SAURON survey was followed by the ATLAS\textsuperscript{3D} survey, which refined the definition of a fast rotating galaxy to take into account the ellipticity of the galaxy, which the original SAURON definition does not.  According to the  ATLAS\textsuperscript{3D} definition, the threshold for a fast rotator is \begin{equation} \lambda_{Re} > (0.31 \pm 0.01) \times \sqrt{\epsilon_e} \end{equation} \citep{Emsellem:11}, where $\epsilon_{e}$ is the ellipticity at the effective radius ($R_e$).  Epsilon is measured by the IDL routine find\_galaxy.pro written by Michele Cappellari and available as part of the mge\_fit\_sectors package\footnotemark[2].

\footnotetext[2]{http://www-astro.physics.ox.ac.uk/$\sim$mxc/idl/}

\begin{figure}
\hspace{ -10 mm }
\epsfig{ file=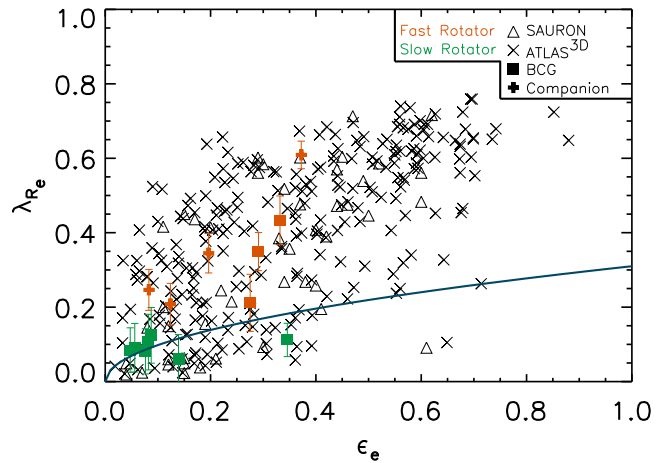, scale=0.55 }
\caption{Lambda at the effective radius as a function of ellipticity, also measured at the effective radius.  SAURON galaxies are plotted as triangles, ATLAS\textsuperscript{3D} galaxies are plotted as crosses, BCGs from this study are plotted as squares, and Companion galaxies from this study are plotted as plus symbols.  The blue line indicates the division between fast rotating and slow rotating galaxies.  We find that three BCGs (1048, 1153, 1261) and four companions (1027, 1066, 1048) are classified as fast rotators.}
\label{lambda_v_e}
\end{figure}

The \lambdare\ vs. $\epsilon_{e}$ plot in Figure \ref{lambda_v_e} shows our quantitative determination of galaxy rotation.  Using the ATLAS\textsuperscript{3D} definition, all galaxies above the blue line are classified as fast rotators.  We find that 3 BCGs and all 4 companions are fast rotators.  Seven of our BCGs are within 1 standard deviation of the dividing line, causing us to doubt their classification.  For them we consider their \lambdar\ profiles to see if they are concave or convex in order to make our determinations.  By the curve criteria, we consider 6 of these ambiguous galaxies to be slow rotators.  Giving us a final total of 3 fast rotating BCGs (30\%) and 4 fast rotating companion galaxies (100\%).

\subsection{Dynamical Mass}

Next we examine the relation between a galaxy's dynamical mass and its \lambdare\ measurement.  We determine the dynamical mass of the galaxy by using the equation provided by \citet{Cappellari:06}:
\begin{equation} M_{dyn} = \frac{5R_e\sigma_e^2}{G} \label{mdyn} \end{equation}
$R_e$ is the effective radius, $\sigma_e$ is the aperture corrected velocity dispersion at the effective radius, and G is the gravitational constant.  The factor of five is a parameter that scales between the virial and Schwarzschild M/L estimates. According to theory it should be 5.953, however those calculations assume perfect one-component isotropic spherical systems, whereas 5 is found to be a good fit for observed galaxies \citep{Cappellari:06}.  In order to compensate for the fact that most of our velocity dispersions reported are measured at $R < R_e$, we have applied the aperture correction from Eq. 1 of \citet{Cappellari:06} to our velocity dispersion results.

\begin{figure}
\hspace{ -5 mm }\epsfig{ file=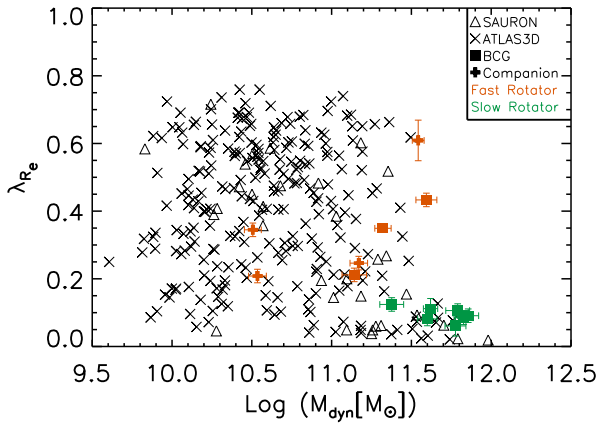, scale=0.5 }
\caption{Angular momentum parameter, \lambdare\ versus dynamical mass, $M_{dyn}$.  SAURON galaxies are plotted as triangles, ATLAS\textsuperscript{3D} galaxies are plotted as crosses, BCGs from this study are plotted as squares, and Companion galaxies from this study are plotted as plus symbols.  The galaxies presented here are amongst the most massive studied.  An upper mass limit to fast rotating galaxies can be seen in all samples around approximately $M_{dyn} = 10^{11.5} M_\sun$.}
\label{lambda_v_mass}
\end{figure}

Dynamical mass results for each galaxy are listed in Table \ref{results_table}.  In Figure \ref{lambda_v_mass} we have plotted \lambdare\ as a function of dynamical mass.  We also plot dynamical masses from the SAURON survey \citep{Emsellem:07} and derived dynamical masses from the reported \atlas\ velocity dispersions \citep{Cappellari:13}.  Figure \ref{lambda_v_mass} shows that the majority of the galaxies reported in this study fall on the high mass end of the ATLAS\textsuperscript{3D} survey sample.  Eight galaxies in our sample have dynamical masses above $10^{11.5} M_{\sun}$.  In the SAURON and \atlas\ results there appears to be a ceiling at approximately $M_{dyn} = 10^{11.5} M_\sun$ at which point galaxies were much less likely to be classified as fast rotators (3/13 (23\%) in \citet{Emsellem:11}).  In agreement with the findings of the ATLAS\textsuperscript{3D} survey, we find that only 25\% of galaxies in our sample above $10^{11.5} M_\sun$ are classified as fast rotating.

\begin{deluxetable*}{ l r r r r r r r}
\tablecolumns{7}
\tablecaption{Kinematic properties of BCGs and companions.  }
\startdata
\hline
\hline
	Galaxy & $R_e$ ($^{\prime\prime}$) & Redshift  & $\sigma_e$ (km s$^{-1}$) & Log(M) (Log($M_{\sun}$)) & $\epsilon$ & \lambdare & FR/SR \\
	 & $\pm$ 0.01 & $\pm$ 0.0001 &  & $\pm$ 0.01 & $\pm$ 0.01 & &  \\
	 
\hline
\\

1027 BCG* &   6.98 & 0.0900 & 227 $\pm$  7 &  11.79 &   0.08 &   0.12 $\pm$  0.08 & SR  \\
1027 Comp* &   4.39 & 0.0909 & 190 $\pm$ 6 & 11.17 &   0.08 &   0.24 $\pm$  0.06 & FR  \\
1042 BCG &   7.22 & 0.0947 & 227 $\pm$  7 & 11.83 &   0.05 &   0.08 $\pm$  0.06 & SR   \\
1048 (A4)\tablenotemark{1} BCG* & 5.17 (3.40)\tablenotemark{1} & 0.0774 & 319 $\pm$ 7  & 11.59 &   0.33 &  0.46 $\pm$ 0.06 & FR  \\
1048 Comp &   1.08 & 0.0801 & 167 $\pm$  3 &  10.51 &   0.20 &   0.34 $\pm$  0.05 & FR  \\
1048 Comp &   1.24 & 0.0746 & 124 $\pm$  3 &  10.54 &   0.12 &   0.21 $\pm$  0.06 & FR  \\
1050 BCG &   8.43 & 0.0722 & 291 $\pm$  11 &  11.78 &   0.14 &   0.07 $\pm$  0.07 & SR   \\
1066 BCG* &   5.07 & 0.0838 & 186 $\pm$  9 &  11.62 &   0.35 &   0.12 $\pm$  0.05 & SR   \\
1066 Comp* & 11.25 & 0.0836 & 176 $\pm$ 13 & 11.55 &   0.37 & 0.60 $\pm$  0.04 & FR  \\
1153 BCG &   2.39 & 0.0591 & 226 $\pm$  7 &  11.14 &   0.28 &   0.21 $\pm$  0.08 & FR  \\
1261 BCG &   5.76 & 0.0371 & 271 $\pm$  1 &  11.32 &   0.29 &   0.35 $\pm$  0.05 & FR  \\
2001 BCG &   5.84 & 0.0415 & 200 $\pm$  4 &  11.38 &   0.09 &   0.13 $\pm$  0.07 & SR   \\
2039 BCG &   8.82 & 0.0829 & 248 $\pm$  6 &   11.86 &   0.06 &   0.10 $\pm$  0.07 & SR   \\
2086 BCG* &  4.83 & 0.0839 & 203 $\pm$  7 &   11.60 &   0.08 &   0.09 $\pm$  0.06 & SR   \\
\enddata
\tablecomments{ Kinematic results from a combination of IDL routines.  $R_e$ is derived from photometric results, but listed here as it is relevant to many values calculated in this table.  (*) Galaxies with an asterisk next to their name are gravitationally bound to their neighboring galaxy.} \tablenotetext{1}{ The effective radius reported in parenthesis for BCG 1048 is for component A4 only, and not the entire BCG.}
\label{results_table}
\end{deluxetable*}

\subsection{Photometry}
\label{photometry}

\begin{figure*}
\epsfig{ file=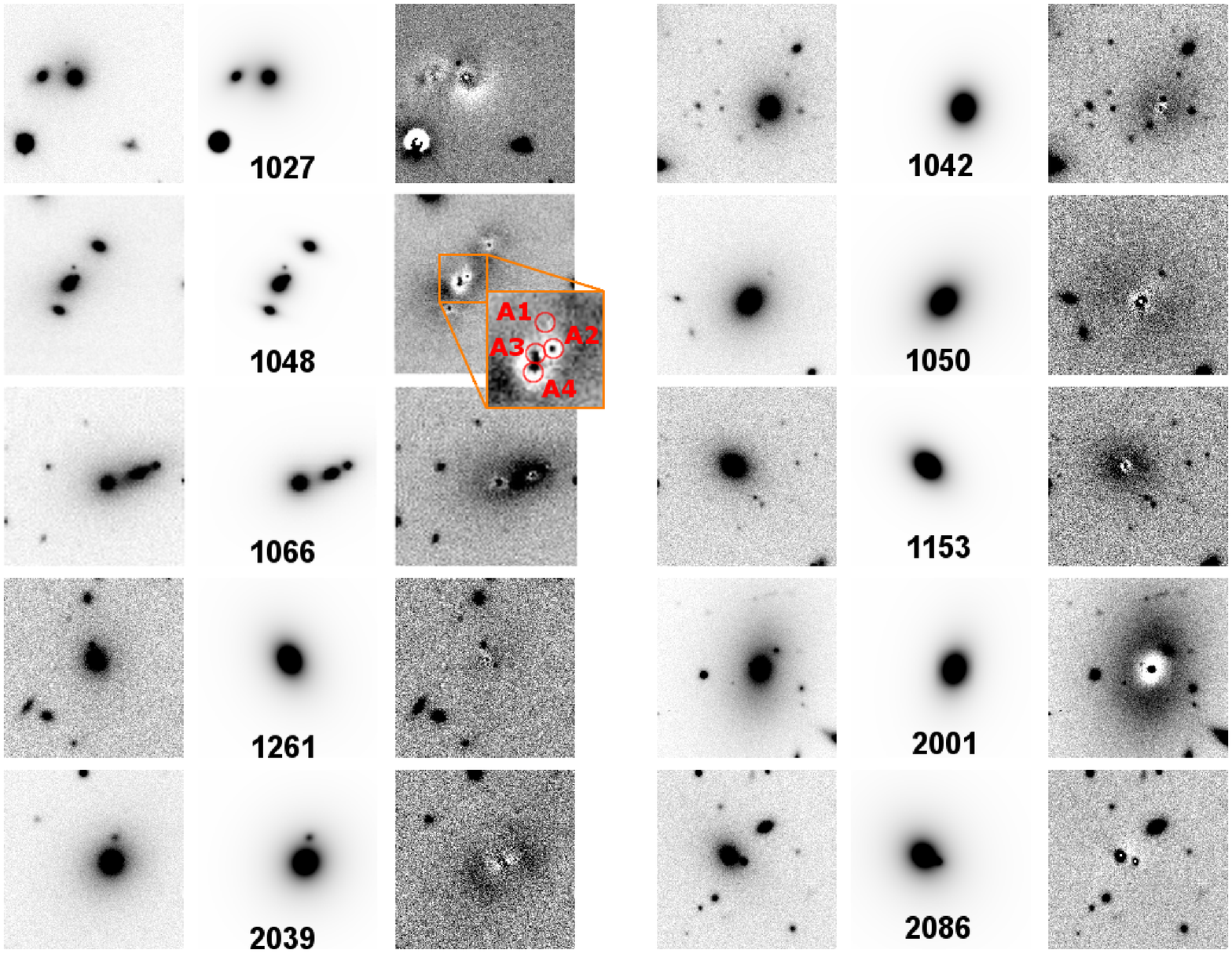, scale = 0.68}
\caption{Single component de Vaucouleurs fits.  Photometric SDSS r-band images of each galaxy is shown on the left hand side for each set.  The middle image is the best fitting de Vaucouleurs model produced by GALFIT, and the right hand side of each set shows the residuals after the GALFIT model has been subtracted from each galaxy.  Each image is $60^{\prime\prime}$ x $60^{\prime\prime}$.  BCG 1048 (FR) is shown modeled by 4 de Vaucouleurs profiles labeled A1-A4 with the center of each profile indicated by the red circles.  Excess light can be seen between components A3 and A4.  Excess light can also be seen in the outskirts of BCGs 1027 (SR), 1048 (FR), 1066 (SR), and 2001 (SR), suggesting on-going or very recent merger event.}
\label{deVaucouleurs}
\end{figure*}

In order to perform our analysis of the spectroscopic data, we have to also examine the photometric data in order to determine the effective radius of each galaxy.  The single component de Vaucouleurs models produced by GALFITS and residual images can be seen in Figure \ref{deVaucouleurs}.   In Figure \ref{deVaucouleurs} one can also see the tidal tails (BCG 1027) and multiple cores (BCG 1048) which motivated us to search for a quantifiable way to measure recent and ongoing mergers.  We find that a single de Vaucouleurs profile is unable to properly model BCG 1048, and the residual image shows multiple hidden cores.  To properly model BCG 1048, we instead use 4 spatially separated de Vaucouleurs profiles.  We report both the single de Vaucouleurs results as well as the results for component A4, brightest of the 4 profiles, in parenthesis in Tables \ref{results_table} and \ref{photometry_table}.

\subsubsection{ $G-M_{20}$ Analysis}

\begin{deluxetable*}{ l r r r r r}
\tablecolumns{6}
\tablecaption{Photometric properties of BCGs and companions.  }
\startdata
\hline
\hline
	Galaxy & $R_e$ ($^{\prime\prime}$) & Integrated Magnitude & G & $M_{20}$ & Merging? \\
	 & $\pm$ 0.01 & $\pm$ 0.01 & $\pm$ 0.01 & $\pm$ 0.01  &  \\
	 
\hline
\\

1027 BCG* &  6.98  & 16.11 & 0.568 & -1.80 & n \\
1027 Comp* &  4.39 & 17.38  & 0.574 & -1.69 & y  \\
1042 BCG & 7.22   & 16.47   & 0.602 & -2.26 & n    \\
1048 (A4)\tablenotemark{1} BCG* & 5.17 (3.40)\tablenotemark{1} & 16.47 (17.07)\tablenotemark{1} & 0.609 & -1.85 & y   \\
1048 Comp &  1.08  & 18.10  & 0.593 & -2.11 & n  \\
1048 Comp & 1.24  & 17.60  & 0.604 & -2.03 & n  \\
1050 BCG & 8.43   & 15.31 & 0.616 & -2.40 & n   \\
1066 BCG* & 4.95  & 16.95 & 0.617 & -2.21 & n    \\
1066 Comp* & 11.95  & 17.05 & 0.613  & -1.83 & y    \\
1153 BCG & 2.39   &  17.22 & 0.626  & -2.35 & n     \\
1261 BCG & 5.76   &  17.36 & 0.576  & -2.38  & n   \\
2001 BCG & 5.84   &  15.75 & 0.527  & -2.20 & n    \\
2039 BCG & 8.82   &  15.63  & 0.559  & -2.17 & n  \\
2086 BCG* & 4.83 &  16.95  & 0.652  & -1.37  & y \\
\enddata
\tablecomments{Photometric results from Source Extractor, GALFIT, and PyMorph.  The effective radius and integrated magnitude come from the de Vaucouleurs fits, where as the Gini and $M_{20}$ values come from PyMorph.  (*) Galaxies with an asterisk next to their name are gravitationally bound to their neighboring galaxy.} \tablenotetext{1}{The effective radius reported in parenthesis for BCG 1048 is for component A4 only, and not the entire BCG.}
\label{photometry_table}
\end{deluxetable*}

The $G-M_{20}$ analysis searches for irregularities in the distribution of light which correlate to morphological signatures of merging.  Our $G-M_{20}$ analysis is performed using PyMorph.  For a galaxy to be classified as a merger candidate it must have \begin{equation} G > -0.14M_{20}+0.33 \end{equation} \citep{Lotz:08}.  As can be seen in Table \ref{photometry_table}, BCG 1048 (FR) is confirmed to be a merger case by the $G-M_{20}$ selection criteria, as is  BCG 2086 (SR).  The other merging candidates are companions to BCGs 1027 (SR), 1066 (SR), and 2086 (SR).

BCG 1261, our fastest rotating galaxy, shows no signs of merging in the $G-M_{20}$ classification.  According to \citet{Lotz:11}, for our gas-poor galaxies, this suggests no recent mergers within the last 0.2 Gyrs by\ the $G-M_{20}$ classification.  That would suggest that any mergers with this $z=0.04$ galaxy would have had to happen before $z=0.06$.

Therefore, 40\% (4/10) of the systems in our sample show morphological signs of current or very recent mergers.  Based on the original unbiased sample of 625 BCGs from \citet{vonderLinden:07}, we find a lower limit of $0.64 \pm 0.32$\% of BCGs are currently undergoing or have undergone a merger within the last 0.2 Gyrs.  Although this subsample is biased towards merging galaxies and not representative of the entire BCG sample, we consider it significant that it is consistent with simulations of BCGs from \citet{DeLucia:07} that showed a 1.47\% merger rate within the same 0.2 Gyrs time period (assuming a constant merger rate from $z=0.5$ to $z=0$) as well as observational results from \citet{Liu:09} that showed a 0.63\% BCG major merger rate within 0.2 Gyrs (assuming a constant merger rate from $z=0.12$ to $z=0.03$).

\subsection{Boundedness}

By utilizing the information from both the photometric and the kinematic results, we are able to determine whether the companion galaxies are likely to be bound to the BCG.  The criteria for being gravitationally bound is 
\begin{equation} V_rR_p \le 2GM sin^2\alpha\ cos\alpha \label{boundedness} \end{equation}
\citep{Beers:82}.  V$_r$ is the radial velocity offset between the two galaxies, R$_p$ is the projected radial separation between the two galaxies, M is the dynamical mass, G is the gravitational constant, and $\alpha$ is the angular separation between the two galaxies.  By integrating over all possible values of $\alpha$, we can find the probability that a companion galaxy will be gravitationally bound to the BCG.

The same analysis was performed by \citet{Brough:11} for BCGs 1027, 1066, and 2086, in which they found that systems 1027 and 1066 were likely to be bound, but 2086 was not.  In performing our analysis, we find that indeed the companion to BCG 1027 has a 65\% chance of being bound, the companion to BCG 1066 has a 94\% chance of being bound, and the companion to BCG 2086 has a 56\% chance of being bound.  We also find in addition that the two outer companions to BCG 1048 (1048 B and 1048 C) never cross the threshold outlined in Equation \ref{boundedness} and are therefore unlikely to be bound to BCG 1048 A.  As seen in Figure \ref{deVaucouleurs}, BCG 1048 A is comprised of 4 components, the most massive being 1048 A2, A3, and  A4 which together make up the galaxy which we call BCG 1048.  We find that component 1048 A3 has an 80\% chance of being bound to 1048 A4.

\vspace{10 mm}

\section{Discussion}

Since the formation history of BCGs is not completely understood, and is in some places contradictory, this research provides vital clues to solving this mystery.  If BCGs form via mostly dry minor mergers where up to half of their mass is accreted isotropically, then most of the initial angular momentum will be removed and we would not expect to see {very massive} BCGs with high values of \lambdare .  If however there are many major merger events with similar orbital parameters to cause a higher angular momentum in the newly formed galaxy despite the lack of gas, then large scale rotation could still be present today.  A third possibility is that there is no connection between merger history and a galaxy's classification as a fast or slow rotator, in which case we would have to look for an outside cause to remove angular momentum.  

\subsection{Mergers and Rotation}

\begin{figure}
\hspace{ -5 mm }\epsfig{ file=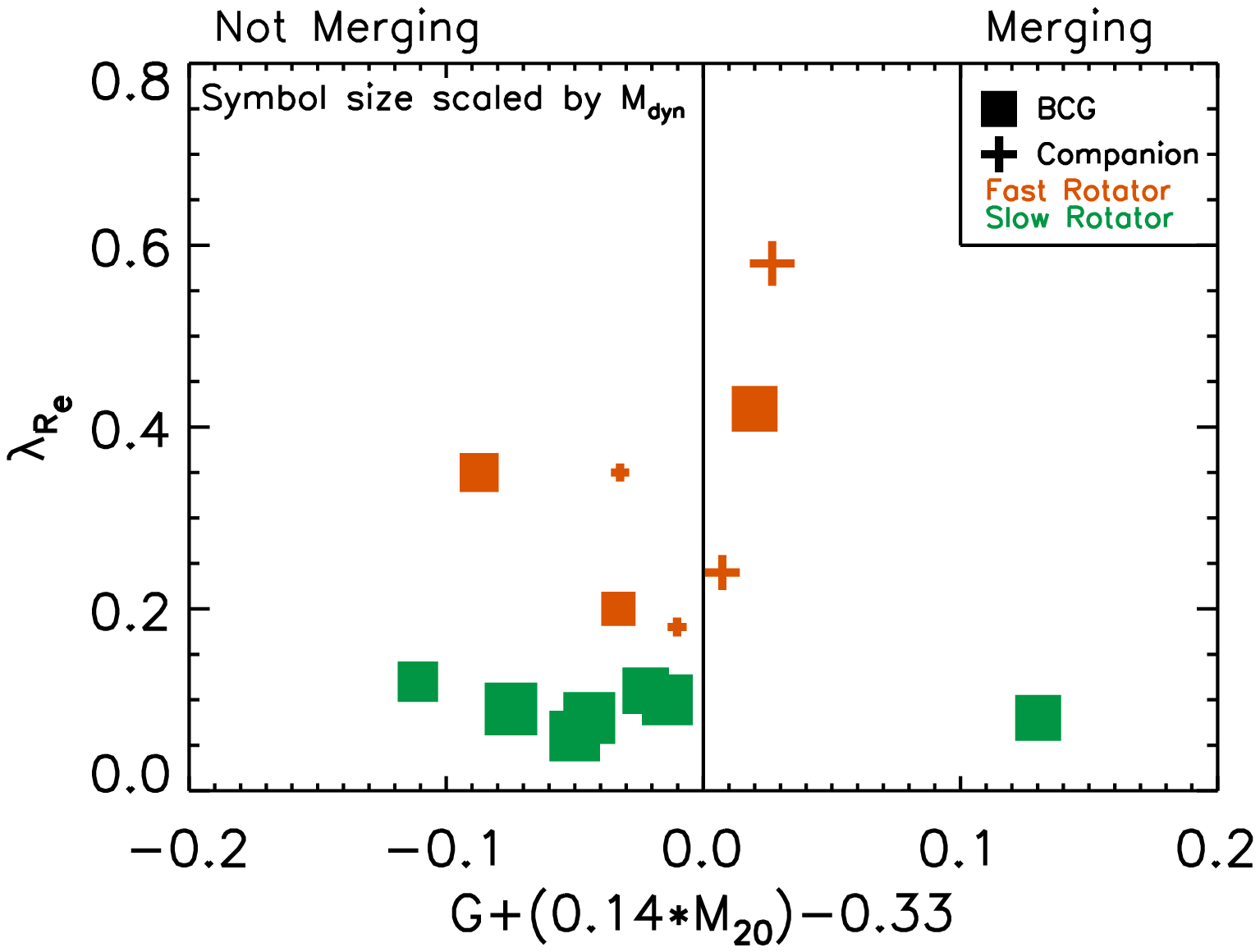, scale=0.5 }
\caption{Merger status and angular momentum at the effective radius.  BCGs are plotted as squares, and companions are plotted as crosses.  Plot symbols are scaled according to galaxy mass, with a larger symbol indicating a higher dynamical mass.  Plot symbols are also color coded, with orange indicating a fast rotating galaxy, and green indicating a slow rotating galaxy.  Galaxies plotted on the negative side of the x-axis are classified as not merging by the $G-M_{20}$ criteria, galaxies on the positive side of the x-axis are classified as merging.  There appears to be no correlation between Merging and \lambdare .}
\label{gini_v_lambdare}
\end{figure}

From the combination of kinematic and photometric results, it appears that BCGs have diverse recent merging histories.  This is unexpected considering their similarities in luminosities and stellar population.  BCGs 1261, 1048, and 1153 are our highest \lambdare\ BCGs and it appears that their high \lambdare\ values come from different sources.  Simulations show that very gas-rich major mergers (\citealt{Cox:06}, \citealt{Robertson:06}) and certain gas-poor major mergers \citep{Bois:11} produce fast rotating galaxies.  Since we do not see any evidence for gas emission lines in any of the galaxies in our sample, we assume that they are still gas-poor.  Below we will discuss specific examples of the various combinations of merging and angular momentum, demonstrating that there is little correlation between the two observables (Figure \ref{gini_v_lambdare}).

\subsubsection{Fast Rotators with No Mergers}

BCGs 1261 \& 1153 are both fast rotators with no evidence of recent mergers.  The photometry for these fast rotating galaxies as measured by $G-M_{20}$ indicates a quiet recent merger history, i.e. no major dry merger events within the last $\sim$ 0.2 Gyrs \citep{Lotz:11}, and no currently occurring mergers.  With our limited timescale we do not have enough information to determine how these fast rotating but very massive BCGs formed.  An analysis of these merger quiet fast rotating BCGs' stellar population would determine whether or not these BCGs are indeed dominated by old stellar populations.

\subsubsection{Fast Rotators with Mergers}

The counterpoint to our fast rotating, quiet history BCGs is BCG 1048 (FR), which is currently undergoing a minor merger.  The multiple cores (A1-A4) present in BCG 1048 (Figure \ref{deVaucouleurs}) suggest an ongoing dry minor merger event, which is coincident with the high \lambdare\ value we measure.  This is further supported by the G-M$_{20}$ selection of BCG 1048 A1-4 as a merging system and the high likelihood that the system is bound.  It is uncertain whether or not the high \lambdare\ value observed in BCG 1048 is the result of actual stellar rotation, or merely the result of having a slightly blue shifted core located near a slightly redshifted core as we do not see the same clear velocity gradient as observed in BCG 1261 (FR).

In order to properly model BCG 1048 (FR) in GALFIT, we require 4 spatially separated components.  By summing the light coming from each model component and assuming that the M/L ratio is the same across the entire BCG, we find evidence for a minor 1:8 merger taking place between A3 and A4.  The other two inner companions, A1 and A2, have mass ratios of approximately 1:8 (A2) and 1:30 (A1) as compared to the central brightest core (A4).  We also find that neither of the outer companion galaxies cross the threshold to be bound to the central merging BCG system for any value of $\alpha$, and they are unlikely to merge in the future.  Despite the fact that the two outermost companions will not merge, the excess light between components A2, A3, and A4 in the central core still shows that this system is actively merging and accreting matter.

For close merging galaxies such as BCG 1048, it is reasonable to question whether the observed kinematic properties, such as velocity dispersion, are artificially inflated by the ongoing merger.  Considering this complication, it is likely that determinations of \lambdare\ for merging systems would not be a robust quantity, and that it may change throughout the merging process.  The measured ellipticity is also likely to be affected during the merging process, causing the BCG to appear to be more elliptical because we are observing two galaxies superimposed upon each other.  However artificially inflating the ellipticity would only serve to increase the threshold for it to be considered a fast rotating galaxy, and would not accidentally classify a slow rotating galaxy as a fast rotating galaxy.

\subsubsection{Slow Rotators without Mergers}

Our slow rotating galaxies have a similarly mixed merging history.  We see that BCGs 1042, 1050, 2001, and 2039 are slow rotators with no evidence of a recent merging event in the $G-M_{20}$ classification, although BCGs 2001 (SR) and 2039 (SR) show a significant amount of enhanced intra-cluster light.  As our results are limited to within $1 R_e$, we are uncertain if it is just the core that is slowly rotating, or if the outer halo is similarly slowly rotating.  IFU measurements of the extended discs seen in BCGs 2001 (SR) and 2039 (SR) could help confirm the results of \citet{Bournaud:04} in which they found that mergers tend to redistribute angular momentum to the outer regions of a galaxy.

\subsubsection{Slow Rotators with Mergers}

BCGs 1027, 1066, and 2086 are slow rotating galaxies currently undergoing a minor merging event according to the $G-M_{20}$ criteria.  All three companions were found to have a more than 50\% chance of being bound to the BCG, adding evidence to the possibility of a minor merger in the future.  Although we do not have enough spaxels with S/N over 10 to determine the rotation of the companion galaxy to BCG 2086, both BCGs 1027 and 1066 are slow rotating BCGs with a fast rotating companions.  BCG 1027 (FR) exhibits tidal tails in the residual photometry, a slight velocity gradient seen in the BCG, and more extreme rotation seen in the companion.  The companion to BCG 1066 (SR) is also a fast rotating galaxy, and appears under visual inspection of the residuals to have two cores, much like the case of BCG 1048 (FR).

\subsection {Companion Rotation}

Companions of BCGs 1066, 1027, and 1048 show clear signs of rotation, both visually and in their \lambdar\ results.  Every BCG companion that we have sufficient data to measure \lambdar\ is a fast rotating galaxy.  Although we have a limited number of spaxels to measure the rotation in BCG 1066, the analysis presented in \citet{Brough:11} provides more spaxels, showing that both companions of 1066 and 1027 are fast rotators.  Our data indicates that all elliptical galaxies near BCGs are fast rotators, however we have an admittedly small sample size of only 4 companion galaxies to make that determination.

\subsection{Angular Momentum and Dynamical Mass}

The presence of fast rotating BCGs in our sample was quite unexpected considering our initial assumption that BCGs should be mostly dispersion supported.  Although, when comparing our measurements of \lambdare\ and $M_{dyn}$ in BCGs to the SAURON and ATLAS\textsuperscript{3D} samples of early-type galaxies (Figure \ref{lambda_v_mass}), we find that our sample is generally consistent with their results.  There appears to be a limit around M$_{dyn} = 10^{11.5} M_{\sun}$ above which there are virtually no galaxies with a high angular momentum, independent of whether they are BCGs or not.  The two galaxies that we do find slightly above this limit are both within 1$\sigma$, and both currently in the process of merging, causing some doubt that they are actually as massive as measured.

In observing these very massive galaxies, in extreme environments, at distances further than what had been measured before, our results suggest that \lambdare\ would be influenced more by a galaxy's dynamical mass than its location within a cluster, and that this is consistent across samples of galaxies that were selected using differing criteria.  The SAURON and ATLAS\textsuperscript{3D} surveys focused on early-type galaxies in general and measured only one BCG, whereas we examined a sample of 10 BCGs at higher redshift.  Larger galaxies likely formed as a result of more mergers than smaller galaxies, therefore they have a higher likelihood of losing their initial angular momentum.  \citet{Peng:10} have also shown that above $M_* = 10^{11} M_{\sun}$ there is a change in the ratio of galaxy mass assembled via post quenched galaxy mergers, also suggesting that these dry merging events could be responsible for removing the angular momentum.

Conversely, \citet{Martizzi:12} have used simulations to show that AGN are a possible mechanism for removing angular momentum from large galaxies.  Having shown in our sample examples of dry mergers that exhibit a high \lambdare\ measurement, it is possible that a different process, such as AGN, are responsible for turning fast rotating galaxies into slow rotators.  It would be interesting and relatively straightforward to test for the presence of the M$_{dyn} = 10^{11.5} M_{\sun}$ limit for fast rotators observed within these samples using such simulations, and to see if the presence of AGN in simulations has any effect on the reproducibility of this limit.

In a further study using higher resolution photometry, we would like to measure which galaxies are cuspy and which are cored as in \citet{Lauer:12}.  It is possible that AGN are responsible for both building the core, as well as removing angular momentum at the same time.  Initial results suggest that our sample of galaxies follow the same relationship seen in the ATLAS\textsuperscript{3D} sample, in that all cored galaxies are slow rotators, but not all slow rotators are cored.  However higher resolution photometry than the SDSS imaging is needed in order to properly make the cuspy/cored determination.

Also in \citet{Lauer:12} they discuss the possibility that the division between fast and slow rotators should be set at a constant \lambdare $_{/2}$ = 0.25.  Using this alternative selection criteria, we would find 2 fast rotating BCGs (20\%), and 2 fast rotating companion galaxies (50\%).  Under this higher threshold, the main message of our results would change little.  The majority of the BCGs that we studied would be slow rotating, however we would still find fast rotating BCGs.  Similarly, 25\% of the galaxies we studied above M$_{dyn} = 10^{11.5} M_{\sun}$ would continue to be fast rotators.

Another possible issue to consider is that in some cases, such as BCG 1048 and the companion to BCG 1066, the galaxies contain multiple cores from ongoing mergers, which could cause the \lambdare\ measurements to be artificially inflated.  This would also artificially increase our measurements of M$_{dyn}$.  It is also possible that our measurements for \lambdare\ for these galaxies will shift as they become dynamically relaxed post-merger.  An examination of \lambdare\ values pre-merger, peri-merger, and post-merger in simulations may shed light on how much of an effect dry mergers would have on \lambdare\ results.  In our analysis, we consider these merging systems, such as BCG 1048 and Companion 1066 to be one galaxy because they are currently in the process of merging.
\subsection{Cluster Rotation}

Another factor to consider would be the angular momentum of the cluster as a whole.  If the angular momentum of a BCG at the kinematic center of a cluster coincides with the average angular momentum of the entire cluster, then infalling galaxies merging with the BCG would likely contribute to the angular momentum of the BCG instead of subtracting from it.  It would be interesting to study the angular momentum of the whole cluster of our best example of a fast rotating BCGs, such as 1261, to see if the angular momentum of the cluster aligns with the angular momentum of the BCG.  Such measurements would require new observations with a much wider field-of-view.  An instrument such as SAMI \citep{Croom:12} with deployable fiber bundles would allow us to collect IFU observations of spatially separated objects across a much wider field of view, e.g. to observations beyond $1 R_e$ for many of the galaxies in our sample.
\\

We have observed that BCGs are surprisingly diverse in their photometric and kinematic properties.  We intend to continue our study of these BCGs by conducting an analysis of their stellar populations in order to further understand their merger history.

\vspace{10 mm}

\section{Conclusion}

In this study we present 2D kinematic and photometric maps of Brightest Cluster Galaxies, ranging in redshift from $z=0.04$ to $z=0.09$ (Figures \ref{single_velocities}, \ref{multi_velocities}, and \ref{deVaucouleurs}).    BCGs are expected to undergo more dry minor mergers than a typical elliptical galaxy, and dry minor mergers have been shown in simulations to decrease a galaxy's angular momentum (\citealt{Bournaud:07}, \citealt{Jesseit:09}, \citealt{Bois:10}, \citealt{Bois:11}).  As the result of many dry minor mergers, we do not expect BCGs to be fast rotating galaxies (as defined by ATLAS\textsuperscript{3D}; \citealt{Emsellem:11}).  We use our sample of 10 BCGs (selected from the SDSS-C4-DR3 catalogue \citep{Miller:05} of BCGs with companions within $10^{\prime\prime}$) and their companions to search for a connection between angular momentum, dynamical mass, and very recent merger history.

Using observations from the VIMOS IFU spectrograph \citep{LeFevre:03} on the VLT we calculate the \lambdar\ values that act as a proxy for angular momentum \citep{Emsellem:07} (Figure \ref{lambda_v_rad}).  We measure \lambdare , \lambdar\ at $1 R_e$, and find \lambdare\ values ranging from $0.07 \pm 0.07$ to $0.60 \pm 0.04$ out of a possible range 0.0 to 1.0 (Figure \ref{lambda_v_e}).  We have shown that 30\% (3/10) of BCGs show signs of galaxy scale rotation and fit the ATLAS\textsuperscript{3D} definition of a fast rotating galaxy and the other 70\% of our BCGs are slow rotating galaxies.  We find the rather surprising result that 100\% (4/4) of our companion galaxies are fast rotating.

Using velocity dispersion (Figure \ref{dispersions}) derived from Equation \ref{mdyn}, we calculate $M_{dyn}$ for our sample of galaxies, finding dynamical masses up to $M_{dyn}  = 10^{11.9} M_\sun$.  BCGs are some of the most massive galaxies in the universe, and the BCGs in our sample are among the most massive galaxies to have their angular momentum calculated using \lambdar\ (Figure \ref{lambda_v_mass}).

While we do find a small number of fast rotating BCGs (Figure \ref{lambda_v_mass}), we find fewer fast rotating galaxies above $M_{dyn}  = 10^{11.5} M_\sun$, consistent with the previous limit for fast rotating galaxies in the SAURON and \atlas\ sample  (\citealt{Emsellem:07}, \citealt{Cappellari:13}).  There appears to be an upper limit to how massive a galaxy can be while still maintaining a high angular momentum.  The consistency between our results and that of the ATLAS\textsuperscript{3D} survey suggests that a galaxy's mass may be more significant than its position inside a cluster halo in determining angular momentum.

Next we examine the ongoing and recent merging history of the galaxies in our sample using the $G-M_{20}$ merger selection criteria \citep{Lotz:08}.  We see that 40\% (4/10) of systems in our sample are currently undergoing or have undergone a merger within the last 0.2 Gyrs (Figure \ref{gini_v_lambdare}).  Although our sample selection is biased towards BCGs that have companions, based on the original unbiased sample of 625 BCGs \citep{vonderLinden:07}, we find that at least $0.64 \pm 0.32$\% of BCGs are undergoing a merger, consistent with observations of BCGs by \citet{Liu:09}.  We use these merging results to examine the connection between \lambdare\ and $G-M_{20}$.

We find no correlation between \lambdare\ and $G-M_{20}$.  We find in 1 BCG that the minor merger event is likely responsible for the high \lambdare\ value observed in the BCG and its companions.  Conversely we find 2 BCGs that are fast rotating without any recent or ongoing merger events.  In the slow rotating BCG category we find 3 mergers connected to a slow rotating BCG, and 4 examples of slow rotating BCGs without any mergers.

These diverse characteristics suggest that there is no strong connection between recent dry mergers and the \lambdare\ classification as a fast or slow rotator.  It is possible that complications from an unrelaxed dynamical system have skewed \lambdare\ measurements in merging systems, or that the removal of angular momentum is due to some other physical process (such as AGN; \citealt{Martizzi:12}) and not predominately due to mergers.  Increased resolution photometry would allow us to measure the cores of these galaxies to see if they are either cusp or cored in order to test the hypothesis of \citet{Lauer:12}.

These surprising results provide clues as to the formation history of Brightest Cluster Galaxies.  It would be beneficial to have an increased sample of BCGs to add to our results, especially BCGs with companions, and more measurements of the companions to determine the probability of a bound BCG companion being a fast rotator.  A larger sample size would also allow us to investigate the possible limit above which fast rotating galaxies do not exist.  Further studies of the ages of the stellar population of seemingly merger quiet galaxies would assist us in understanding their merger histories beyond 0.2 Gyrs.  Our observations lead us to conclude that there is much diversity in the measurements of a BCG's angular momentum, dynamical mass, and merger history (Figure \ref{gini_v_lambdare}).

\vspace{10 mm}

\section{Acknowledgments}
We would like to thank the VIMOS team for their pipeline, and GitHub for hosting the code used to reduce the data.  We would also like to thank Brett Salmon and Adam Tomczack for their assistance throughout the process of writing this paper.  We would also like to thank Louis Abramson for assistance with GALFIT and Source Extractor.  We would like to thank our referee Eric Emsellem for his excellent feedback, which helped to clarify our narrative as well as our results.

The data reduction performed to obtain our results was done on the Brazos supercomputing cluster at Texas A\&M university.  We would especially like to thank the Mitchell family for their continuing support and in particular the late George P. Mitchell whose vision and commitment to science and astronomy leaves a lasting legacy.

\bibliographystyle{apj}
\bibliography{Jimmy2013}

\begin{thebibliography}{54}
\expandafter\ifx\csname natexlab\endcsname\relax\def\natexlab#1{#1}\fi

\bibitem[{{Abraham} {et~al.}(2003){Abraham}, {van den Bergh}, \&
  {Nair}}]{Abraham:03}
{Abraham}, R.~G., {van den Bergh}, S., \& {Nair}, P. 2003, \apj, 588, 218

\bibitem[{{Beers} {et~al.}(1982){Beers}, {Geller}, \& {Huchra}}]{Beers:82}
{Beers}, T.~C., {Geller}, M.~J., \& {Huchra}, J.~P. 1982, \apj, 257, 23

\bibitem[{{Bender} {et~al.}(1992){Bender}, {Burstein}, \& {Faber}}]{Bender:92}
{Bender}, R., {Burstein}, D., \& {Faber}, S.~M. 1992, \apj, 399, 462

\bibitem[{{Bernardi}(2009)}]{Bernardi:09}
{Bernardi}, M. 2009, \mnras, 395, 1491

\bibitem[{{Bertin} \& {Arnouts}(1996)}]{Bertin:96}
{Bertin}, E., \& {Arnouts}, S. 1996, \aaps, 117, 393

\bibitem[{{Bois} {et~al.}(2010){Bois}, {Bournaud}, {Emsellem}, {Alatalo},
  {Blitz}, {Bureau}, {Cappellari}, {Davies}, {Davis}, {de Zeeuw}, {Duc},
  {Khochfar}, {Krajnovi{\'c}}, {Kuntschner}, {Lablanche}, {McDermid},
  {Morganti}, {Naab}, {Oosterloo}, {Sarzi}, {Scott}, {Serra}, {Weijmans}, \&
  {Young}}]{Bois:10}
{Bois}, M., {Bournaud}, F., {Emsellem}, E., {et~al.} 2010, \mnras, 406, 2405

\bibitem[{{Bois} {et~al.}(2011){Bois}, {Emsellem}, {Bournaud}, {Alatalo},
  {Blitz}, {Bureau}, {Cappellari}, {Davies}, {Davis}, {de Zeeuw}, {Duc},
  {Khochfar}, {Krajnovi{\'c}}, {Kuntschner}, {Lablanche}, {McDermid},
  {Morganti}, {Naab}, {Oosterloo}, {Sarzi}, {Scott}, {Serra}, {Weijmans}, \&
  {Young}}]{Bois:11}
{Bois}, M., {Emsellem}, E., {Bournaud}, F., {et~al.} 2011, \mnras, 416, 1654

\bibitem[{{Bournaud} {et~al.}(2004){Bournaud}, {Combes}, \&
  {Jog}}]{Bournaud:04}
{Bournaud}, F., {Combes}, F., \& {Jog}, C.~J. 2004, \aap, 418, L27

\bibitem[{{Bournaud} {et~al.}(2007){Bournaud}, {Jog}, \&
  {Combes}}]{Bournaud:07}
{Bournaud}, F., {Jog}, C.~J., \& {Combes}, F. 2007, \aap, 476, 1179

\bibitem[{{Boylan-Kolchin} {et~al.}(2006){Boylan-Kolchin}, {Ma}, \&
  {Quataert}}]{Boylan-Kolchin:08}
{Boylan-Kolchin}, M., {Ma}, C.-P., \& {Quataert}, E. 2006, \mnras, 369, 1081

\bibitem[{{Brough} {et~al.}(2005){Brough}, {Collins}, {Burke}, {Lynam}, \&
  {Mann}}]{Brough:05}
{Brough}, S., {Collins}, C.~A., {Burke}, D.~J., {Lynam}, P.~D., \& {Mann},
  R.~G. 2005, \mnras, 364, 1354

\bibitem[{{Brough} {et~al.}(2011){Brough}, {Tran}, {Sharp}, {von der Linden},
  \& {Couch}}]{Brough:11}
{Brough}, S., {Tran}, K.-V., {Sharp}, R.~G., {von der Linden}, A., \& {Couch},
  W.~J. 2011, \mnras, 414, L80

\bibitem[{{Cappellari} \& {Copin}(2003)}]{Cappellari:03}
{Cappellari}, M., \& {Copin}, Y. 2003, \mnras, 342, 345

\bibitem[{{Cappellari} \& {Emsellem}(2004)}]{Cappellari:04}
{Cappellari}, M., \& {Emsellem}, E. 2004, \pasp, 116, 138

\bibitem[{{Cappellari} {et~al.}(2006){Cappellari}, {Bacon}, {Bureau}, {Damen},
  {Davies}, {de Zeeuw}, {Emsellem}, {Falc{\'o}n-Barroso}, {Krajnovi{\'c}},
  {Kuntschner}, {McDermid}, {Peletier}, {Sarzi}, {van den Bosch}, \& {van de
  Ven}}]{Cappellari:06}
{Cappellari}, M., {Bacon}, R., {Bureau}, M., {et~al.} 2006, \mnras, 366, 1126

\bibitem[{{Cappellari} {et~al.}(2013){Cappellari}, {Scott}, {Alatalo}, {Blitz},
  {Bois}, {Bournaud}, {Bureau}, {Crocker}, {Davies}, {Davis}, {de Zeeuw},
  {Duc}, {Emsellem}, {Khochfar}, {Krajnovi{\'c}}, {Kuntschner}, {McDermid},
  {Morganti}, {Naab}, {Oosterloo}, {Sarzi}, {Serra}, {Weijmans}, \&
  {Young}}]{Cappellari:13}
{Cappellari}, M., {Scott}, N., {Alatalo}, K., {et~al.} 2013, \mnras, 432, 1709

\bibitem[{{Cox} {et~al.}(2006){Cox}, {Dutta}, {Di Matteo}, {Hernquist},
  {Hopkins}, {Robertson}, \& {Springel}}]{Cox:06}
{Cox}, T.~J., {Dutta}, S.~N., {Di Matteo}, T., {et~al.} 2006, \apj, 650, 791

\bibitem[{{Croom} {et~al.}(2012){Croom}, {Lawrence}, {Bland-Hawthorn},
  {Bryant}, {Fogarty}, {Richards}, {Goodwin}, {Farrell}, {Miziarski}, {Heald},
  {Jones}, {Lee}, {Colless}, {Brough}, {Hopkins}, {Bauer}, {Birchall}, {Ellis},
  {Horton}, {Leon-Saval}, {Lewis}, {L{\'o}pez-S{\'a}nchez}, {Min}, {Trinh}, \&
  {Trowland}}]{Croom:12}
{Croom}, S.~M., {Lawrence}, J.~S., {Bland-Hawthorn}, J., {et~al.} 2012, \mnras,
  421, 872

\bibitem[{{de Jong} {et~al.}(2004){de Jong}, {Kassin}, {Bell}, \&
  {Courteau}}]{deJong:04}
{de Jong}, R.~S., {Kassin}, S., {Bell}, E.~F., \& {Courteau}, S. 2004, in IAU
  Symposium, Vol. 220, Dark Matter in Galaxies, ed. S.~{Ryder}, D.~{Pisano},
  M.~{Walker}, \& K.~{Freeman}, 281

\bibitem[{{De Lucia} \& {Blaizot}(2007)}]{DeLucia:07}
{De Lucia}, G., \& {Blaizot}, J. 2007, \mnras, 375, 2

\bibitem[{{Desroches} {et~al.}(2007){Desroches}, {Quataert}, {Ma}, \&
  {West}}]{Desroches:07}
{Desroches}, L.-B., {Quataert}, E., {Ma}, C.-P., \& {West}, A.~A. 2007, \mnras,
  377, 402

\bibitem[{{D'Onghia} \& {Burkert}(2004)}]{D'Onghia:04}
{D'Onghia}, E., \& {Burkert}, A. 2004, \apjl, 612, L13

\bibitem[{{Doroshkevich}(1970)}]{Doroshkevich:70}
{Doroshkevich}, A.~G. 1970, Astrophysics, 6, 320

\bibitem[{{Emsellem} {et~al.}(2007){Emsellem}, {Cappellari}, {Krajnovi{\'c}},
  {van de Ven}, {Bacon}, {Bureau}, {Davies}, {de Zeeuw}, {Falc{\'o}n-Barroso},
  {Kuntschner}, {McDermid}, {Peletier}, \& {Sarzi}}]{Emsellem:07}
{Emsellem}, E., {Cappellari}, M., {Krajnovi{\'c}}, D., {et~al.} 2007, \mnras,
  379, 401

\bibitem[{{Emsellem} {et~al.}(2011){Emsellem}, {Cappellari}, {Krajnovi{\'c}},
  {Alatalo}, {Blitz}, {Bois}, {Bournaud}, {Bureau}, {Davies}, {Davis}, {de
  Zeeuw}, {Khochfar}, {Kuntschner}, {Lablanche}, {McDermid}, {Morganti},
  {Naab}, {Oosterloo}, {Sarzi}, {Scott}, {Serra}, {van de Ven}, {Weijmans}, \&
  {Young}}]{Emsellem:11}
---. 2011, \mnras, 414, 888

\bibitem[{{Faber} {et~al.}(1997){Faber}, {Tremaine}, {Ajhar}, {Byun},
  {Dressler}, {Gebhardt}, {Grillmair}, {Kormendy}, {Lauer}, \&
  {Richstone}}]{Faber:97}
{Faber}, S.~M., {Tremaine}, S., {Ajhar}, E.~A., {et~al.} 1997, \aj, 114, 1771

\bibitem[{{Izzo} {et~al.}(2004){Izzo}, {Kornweibel}, {McKay}, {Palsa}, {Peron},
  \& {Taylor}}]{Izzo:04}
{Izzo}, C., {Kornweibel}, N., {McKay}, D., {et~al.} 2004, The Messenger, 117,
  33

\bibitem[{{Jesseit} {et~al.}(2009){Jesseit}, {Cappellari}, {Naab}, {Emsellem},
  \& {Burkert}}]{Jesseit:09}
{Jesseit}, R., {Cappellari}, M., {Naab}, T., {Emsellem}, E., \& {Burkert}, A.
  2009, \mnras, 397, 1202

\bibitem[{{Kormendy} \& {Bender}(1996)}]{Kormendy:96}
{Kormendy}, J., \& {Bender}, R. 1996, \apjl, 464, L119

\bibitem[{{Lauer}(2012)}]{Lauer:12}
{Lauer}, T.~R. 2012, \apj, 759, 64

\bibitem[{{Lauer} {et~al.}(2007){Lauer}, {Faber}, {Richstone}, {Gebhardt},
  {Tremaine}, {Postman}, {Dressler}, {Aller}, {Filippenko}, {Green}, {Ho},
  {Kormendy}, {Magorrian}, \& {Pinkney}}]{Lauer:07}
{Lauer}, T.~R., {Faber}, S.~M., {Richstone}, D., {et~al.} 2007, \apj, 662, 808

\bibitem[{{Le F{\`e}vre} {et~al.}(2003){Le F{\`e}vre}, {Saisse}, {Mancini},
  {Brau-Nogue}, {Caputi}, {Castinel}, {D'Odorico}, {Garilli}, {Kissler-Patig},
  {Lucuix}, {Mancini}, {Pauget}, {Sciarretta}, {Scodeggio}, {Tresse}, \&
  {Vettolani}}]{LeFevre:03}
{Le F{\`e}vre}, O., {Saisse}, M., {Mancini}, D., {et~al.} 2003, in Society of
  Photo-Optical Instrumentation Engineers (SPIE) Conference Series, Vol. 4841,
  Society of Photo-Optical Instrumentation Engineers (SPIE) Conference Series,
  ed. M.~{Iye} \& A.~F.~M. {Moorwood}, 1670--1681

\bibitem[{{Lidman} {et~al.}(2012){Lidman}, {Suherli}, {Muzzin}, {Wilson},
  {Demarco}, {Brough}, {Rettura}, {Cox}, {DeGroot}, {Yee}, {Gilbank},
  {Hoekstra}, {Balogh}, {Ellingson}, {Hicks}, {Nantais}, {Noble}, {Lacy},
  {Surace}, \& {Webb}}]{Lidman:12}
{Lidman}, C., {Suherli}, J., {Muzzin}, A., {et~al.} 2012, ArXiv e-prints

\bibitem[{{Liu} {et~al.}(2009){Liu}, {Mao}, {Deng}, {Xia}, \& {Wen}}]{Liu:09}
{Liu}, F.~S., {Mao}, S., {Deng}, Z.~G., {Xia}, X.~Y., \& {Wen}, Z.~L. 2009,
  \mnras, 396, 2003

\bibitem[{{Lotz} {et~al.}(2011){Lotz}, {Jonsson}, {Cox}, {Croton}, {Primack},
  {Somerville}, \& {Stewart}}]{Lotz:11}
{Lotz}, J.~M., {Jonsson}, P., {Cox}, T.~J., {et~al.} 2011, \apj, 742, 103

\bibitem[{{Lotz} {et~al.}(2008){Lotz}, {Jonsson}, {Cox}, \&
  {Primack}}]{Lotz:08}
{Lotz}, J.~M., {Jonsson}, P., {Cox}, T.~J., \& {Primack}, J.~R. 2008, \mnras,
  391, 1137

\bibitem[{{Martizzi} {et~al.}(2012){Martizzi}, {Teyssier}, \&
  {Moore}}]{Martizzi:12}
{Martizzi}, D., {Teyssier}, R., \& {Moore}, B. 2012, \mnras, 420, 2859

\bibitem[{{Miller} {et~al.}(2005){Miller}, {Nichol}, {Reichart}, {Wechsler},
  {Evrard}, {Annis}, {McKay}, {Bahcall}, {Bernardi}, {Boehringer}, {Connolly},
  {Goto}, {Kniazev}, {Lamb}, {Postman}, {Schneider}, {Sheth}, \&
  {Voges}}]{Miller:05}
{Miller}, C.~J., {Nichol}, R.~C., {Reichart}, D., {et~al.} 2005, \aj, 130, 968

\bibitem[{{Naab} {et~al.}(2006){Naab}, {Jesseit}, \& {Burkert}}]{Naab:06a}
{Naab}, T., {Jesseit}, R., \& {Burkert}, A. 2006, \mnras, 372, 839

\bibitem[{{Oegerle} \& {Hoessel}(1991)}]{Oegerle:91}
{Oegerle}, W.~R., \& {Hoessel}, J.~G. 1991, \apj, 375, 15

\bibitem[{{Peebles}(1969)}]{Peebles:69}
{Peebles}, P.~J.~E. 1969, \apj, 155, 393

\bibitem[{{Peng} {et~al.}(2002){Peng}, {Ho}, {Impey}, \& {Rix}}]{Peng:02}
{Peng}, C.~Y., {Ho}, L.~C., {Impey}, C.~D., \& {Rix}, H.-W. 2002, \aj, 124, 266

\bibitem[{{Peng} {et~al.}(2010){Peng}, {Lilly}, {Kova{\v c}}, {Bolzonella},
  {Pozzetti}, {Renzini}, {Zamorani}, {Ilbert}, {Knobel}, {Iovino}, {Maier},
  {Cucciati}, {Tasca}, {Carollo}, {Silverman}, {Kampczyk}, {de Ravel},
  {Sanders}, {Scoville}, {Contini}, {Mainieri}, {Scodeggio}, {Kneib}, {Le
  F{\`e}vre}, {Bardelli}, {Bongiorno}, {Caputi}, {Coppa}, {de la Torre},
  {Franzetti}, {Garilli}, {Lamareille}, {Le Borgne}, {Le Brun}, {Mignoli},
  {Perez Montero}, {Pello}, {Ricciardelli}, {Tanaka}, {Tresse}, {Vergani},
  {Welikala}, {Zucca}, {Oesch}, {Abbas}, {Barnes}, {Bordoloi}, {Bottini},
  {Cappi}, {Cassata}, {Cimatti}, {Fumana}, {Hasinger}, {Koekemoer},
  {Leauthaud}, {Maccagni}, {Marinoni}, {McCracken}, {Memeo}, {Meneux}, {Nair},
  {Porciani}, {Presotto}, \& {Scaramella}}]{Peng:10}
{Peng}, Y.-j., {Lilly}, S.~J., {Kova{\v c}}, K., {et~al.} 2010, \apj, 721, 193

\bibitem[{{Robertson} {et~al.}(2006){Robertson}, {Cox}, {Hernquist}, {Franx},
  {Hopkins}, {Martini}, \& {Springel}}]{Robertson:06}
{Robertson}, B., {Cox}, T.~J., {Hernquist}, L., {et~al.} 2006, \apj, 641, 21

\bibitem[{{S{\'a}nchez-Bl{\'a}zquez} {et~al.}(2006){S{\'a}nchez-Bl{\'a}zquez},
  {Peletier}, {Jim{\'e}nez-Vicente}, {Cardiel}, {Cenarro},
  {Falc{\'o}n-Barroso}, {Gorgas}, {Selam}, \& {Vazdekis}}]{Sanchez-Blazquez:06}
{S{\'a}nchez-Bl{\'a}zquez}, P., {Peletier}, R.~F., {Jim{\'e}nez-Vicente}, J.,
  {et~al.} 2006, \mnras, 371, 703

\bibitem[{{Stott} {et~al.}(2011){Stott}, {Collins}, {Burke}, {Hamilton-Morris},
  \& {Smith}}]{Stott:11}
{Stott}, J.~P., {Collins}, C.~A., {Burke}, C., {Hamilton-Morris}, V., \&
  {Smith}, G.~P. 2011, \mnras, 414, 445

\bibitem[{{Tacconi} {et~al.}(2010){Tacconi}, {Genzel}, {Neri}, {Cox}, {Cooper},
  {Shapiro}, {Bolatto}, {Bouch{\'e}}, {Bournaud}, {Burkert}, {Combes},
  {Comerford}, {Davis}, {Schreiber}, {Garcia-Burillo}, {Gracia-Carpio}, {Lutz},
  {Naab}, {Omont}, {Shapley}, {Sternberg}, \& {Weiner}}]{Tacconi:10}
{Tacconi}, L.~J., {Genzel}, R., {Neri}, R., {et~al.} 2010, \nat, 463, 781

\bibitem[{{Tran} {et~al.}(2008){Tran}, {Moustakas}, {Gonzalez}, {Bai},
  {Zaritsky}, \& {Kautsch}}]{Tran:08}
{Tran}, K.-V.~H., {Moustakas}, J., {Gonzalez}, A.~H., {et~al.} 2008, \apjl,
  683, L17

\bibitem[{{van den Bosch} {et~al.}(2002){van den Bosch}, {Abel}, {Croft},
  {Hernquist}, \& {White}}]{vandenBosch:02}
{van den Bosch}, F.~C., {Abel}, T., {Croft}, R.~A.~C., {Hernquist}, L., \&
  {White}, S.~D.~M. 2002, \apj, 576, 21

\bibitem[{{Vikram} {et~al.}(2010){Vikram}, {Wadadekar}, {Kembhavi}, \&
  {Vijayagovindan}}]{Vikram:10}
{Vikram}, V., {Wadadekar}, Y., {Kembhavi}, A.~K., \& {Vijayagovindan}, G.~V.
  2010, \mnras, 409, 1379

\bibitem[{{Vitvitska} {et~al.}(2002){Vitvitska}, {Klypin}, {Kravtsov},
  {Wechsler}, {Primack}, \& {Bullock}}]{Vitvitska:02}
{Vitvitska}, M., {Klypin}, A.~A., {Kravtsov}, A.~V., {et~al.} 2002, \apj, 581,
  799

\bibitem[{{von der Linden} {et~al.}(2007){von der Linden}, {Best}, {Kauffmann},
  \& {White}}]{vonderLinden:07}
{von der Linden}, A., {Best}, P.~N., {Kauffmann}, G., \& {White}, S.~D.~M.
  2007, \mnras, 379, 867

\bibitem[{{White}(1984)}]{White:84}
{White}, S.~D.~M. 1984, \apj, 286, 38

\bibitem[{{York} {et~al.}(2000){York}, {Adelman}, {Anderson}, {Anderson},
  {Annis}, {Bahcall}, {Bakken}, {Barkhouser}, {Bastian}, {Berman}, {Boroski},
  {Bracker}, {Briegel}, {Briggs}, {Brinkmann}, {Brunner}, {Burles}, {Carey},
  {Carr}, {Castander}, {Chen}, {Colestock}, {Connolly}, {Crocker}, {Csabai},
  {Czarapata}, {Davis}, {Doi}, {Dombeck}, {Eisenstein}, {Ellman}, {Elms},
  {Evans}, {Fan}, {Federwitz}, {Fiscelli}, {Friedman}, {Frieman}, {Fukugita},
  {Gillespie}, {Gunn}, {Gurbani}, {de Haas}, {Haldeman}, {Harris}, {Hayes},
  {Heckman}, {Hennessy}, {Hindsley}, {Holm}, {Holmgren}, {Huang}, {Hull},
  {Husby}, {Ichikawa}, {Ichikawa}, {Ivezi{\'c}}, {Kent}, {Kim}, {Kinney},
  {Klaene}, {Kleinman}, {Kleinman}, {Knapp}, {Korienek}, {Kron}, {Kunszt},
  {Lamb}, {Lee}, {Leger}, {Limmongkol}, {Lindenmeyer}, {Long}, {Loomis},
  {Loveday}, {Lucinio}, {Lupton}, {MacKinnon}, {Mannery}, {Mantsch}, {Margon},
  {McGehee}, {McKay}, {Meiksin}, {Merelli}, {Monet}, {Munn}, {Narayanan},
  {Nash}, {Neilsen}, {Neswold}, {Newberg}, {Nichol}, {Nicinski}, {Nonino},
  {Okada}, {Okamura}, {Ostriker}, {Owen}, {Pauls}, {Peoples}, {Peterson},
  {Petravick}, {Pier}, {Pope}, {Pordes}, {Prosapio}, {Rechenmacher}, {Quinn},
  {Richards}, {Richmond}, {Rivetta}, {Rockosi}, {Ruthmansdorfer}, {Sandford},
  {Schlegel}, {Schneider}, {Sekiguchi}, {Sergey}, {Shimasaku}, {Siegmund},
  {Smee}, {Smith}, {Snedden}, {Stone}, {Stoughton}, {Strauss}, {Stubbs},
  {SubbaRao}, {Szalay}, {Szapudi}, {Szokoly}, {Thakar}, {Tremonti}, {Tucker},
  {Uomoto}, {Vanden Berk}, {Vogeley}, {Waddell}, {Wang}, {Watanabe},
  {Weinberg}, {Yanny}, {Yasuda}, \& {SDSS Collaboration}}]{York:00}
{York}, D.~G., {Adelman}, J., {Anderson}, Jr., J.~E., {et~al.} 2000, \aj, 120,
  1579

\end{thebibliography}

\end{document}